\documentclass{emulateapj}
\shortauthors{Matthews et al.}
\shorttitle{Evolving Radio Photospheres}

\begin{document}
\newcommand{\ang}{\rm \AA}
\newcommand{\msun}{M$_\odot$}
\newcommand{\lsun}{L$_\odot$}
\newcommand{\days}{$d$}
\newcommand{\degree}{$^\circ$}
\newcommand{\ud}{{\rm d}}
\newcommand{\as}[2]{$#1''\,\hspace{-1.7mm}.\hspace{.0mm}#2$}
\newcommand{\am}[2]{$#1'\,\hspace{-1.7mm}.\hspace{.0mm}#2$}
\newcommand{\ad}[2]{$#1^{\circ}\,\hspace{-1.7mm}.\hspace{.0mm}#2$}
\newcommand{\lsim}{~\rlap{$<$}{\lower 1.0ex\hbox{$\sim$}}}
\newcommand{\gsim}{~\rlap{$>$}{\lower 1.0ex\hbox{$\sim$}}}
\newcommand{\HA}{H$\alpha$}
\newcommand{\HII}{\mbox{H\,{\sc ii}}}
\newcommand{\kms}{\mbox{km s$^{-1}$}}
\newcommand{\HI}{\mbox{H\,{\sc i}}}
\newcommand{\jks}{Jy~km~s$^{-1}$}

\title{The Evolving Radio Photospheres of Long-Period Variable Stars}

\author{L. D. Matthews\altaffilmark{1},  M. J. Reid\altaffilmark{2}, 
K. M. Menten\altaffilmark{3}, K. Akiyama\altaffilmark{1},\altaffilmark{4}}

\altaffiltext{1}{MIT Haystack Observatory, 99 Millstone Road, Westford, MA
  01886 USA}
\altaffiltext{2}{Harvard-Smithsonian Center for Astrophysics, 60
  Garden Street, MS-42, Cambridge, MA 02138 USA}
\altaffiltext{3}{Max-Planck-Institut f\"ur Radioastronomie, Auf dem
  H\"ugel 69, D-53121 Bonn, Germany}
\altaffiltext{4}{National Radio Astronomy Observatory, 520 Edgemont
  Rd, Charlottesville, VA 22903, USA} 

\begin{abstract}
Observations with the Karl G. Jansky Very Large Array at 46~GHz 
($\lambda\approx$7~mm) have been used to measure the size and shape of 
the radio photospheres of four
long-period variable stars: R~Leonis (R~Leo), IRC+10216
(CW~Leo), 
$\chi$~Cygni ($\chi$~Cyg), and  W~Hydrae (W~Hya). The shapes of the stars range
from nearly round to ellipticities of $\sim$0.15. Comparisons with
observations taken several years earlier
show that the photospheric parameters (mean diameter, shape,
and/or flux density) of each of the stars have changed
over time. Evidence for brightness asymmetries and non-uniformities across the radio
surfaces are also seen in the visibility domain
and in images obtained using a sparse modeling image reconstruction
technique. 
These trends may be explained 
as manifestations of large-scale irregular convective flows on the
stellar surface, although effects from non-radial
pulsations cannot be excluded. Our data also allow a new
evaluation
of the proper motion of IRC+10216. Our measurement is in agreement
with previous values obtained from radio wavelength measurements, and
we find no evidence of statistically significant astrometric perturbations from a
binary companion. 

\end{abstract}

\keywords{stars: AGB and
post-AGB -- stars: atmospheres -- stars: fundamental parameters --
stars: imaging}  

\section{Introduction\protect\label{Intro}}
Stars on the asymptotic giant branch (AGB) emit continuum radiation at
centimeter and
(sub)millimeter wavelength from a ``radio photosphere'' with 
a radius approximately twice that of the classical photospheric
radius $R_{\star}$, defined by the line-free regions of the optical-infrared
spectrum (Reid \& Menten 1997; 
hereafter RM97).  For Mira-type variables, $R_{\star}$ is typically
$\sim$1--2~AU. Just outside of this radius, at  $r\sim$1--$2R_{\star}$, there
exists a molecular layer (sometimes referred to as the ``MOLsphere'';
Tsuji 2000, 2001)
that may be nearly opaque in the visible and infrared (Reid \&
Goldston 2002; Perrin et al. 2004; Tsuji 2008). 
The radio photosphere lies near the outskirts of this molecular layer,
at $\sim2~R_{\star}$.

The radio photosphere exists within
a critical juncture between the stellar ``surface''
and the atmospheric regions several AU further out, where the
stellar wind is launched. AGB star winds are generally assumed to be
dust-driven. This requires the transport of gas from the warmer, dust-free regions
of the photosphere to heights cool enough for dust grain formation
and survival. The gas transport is believed to involve pulsations, convection, and/or
shocks (e.g., H\"ofner 2008). Studying the characteristics of the
radio photosphere and its temporal changes
can therefore provide valuable diagnostic information on the relative importance of these different
processes. In cases where radio photospheres can be spatially resolved, such observations
can also be used to help test the increasingly
sophisticated two- and three-dimensional (3D) models of AGB star
atmospheres 
that are becoming available
(Woitke 2006; Freytag \& H\"ofner 2008; Freytag, Liljegren, \&
H\"ofner  2017). 

Spatially resolved imaging observations are possible for AGB stars within
$d\lsim$200~pc
using the most extended configurations of the Very Large Array (VLA)
and the Atacama Large Millimeter/submillimeter Array (ALMA). 
Reid \& Menten (2007; hereafter RM07) 
published the first resolved 7~mm images of the radio photospheres of three
oxygen-rich AGB stars [R~Leo, Mira ($o$~Ceti), and W~Hya] obtained with the
VLA, and Menten et al. (2012; hereafter M12) published similar observations for the
carbon star IRC+10216 (CW~Leo). These observations revealed an intriguing result:
two of the four stars observed (R~Leo and W~Hya) exhibited statistically
significant deviations from sphericity, while Mira and IRC+10216
appeared approximately spherical. However, in 2014 observations
from the VLA and ALMA,  Matthews
et al. (2015) found that Mira too appeared ``squashed'' (see also 
Vlemmings et al. 2015; Wong et al. 2016). These results further hinted that
Mira's shape
has evolved with time.

Deviations from sphericity in AGB stars could have several possible
causes, either intrinsic or extrinsic. These include rotational
flattening (e.g., if the star were spun up by mass accretion or an
in-spiraling companion; e.g., Livio 1994), tidal
forces from a companion (e.g., Huggins et al. 1990), 
magnetic effects (e.g., Szymczak et al. 1998), non-radial pulsations
(e.g., Tuthill et al. 1994; Stello et al. 2014; Wood 2015), or
manifestations of 
irregular convective flows resulting from the interactions between
large-scale convective cells and the stellar pulsation (Freytag et al. 2017).
Crucial to distinguishing between these possibilities is determining
whether the observed shapes are static or variable.

To explore this topic further, we present here new 7~mm imaging
observations of four nearby AGB stars, including the three stars previously
imaged by MR07
and M12. Our study includes the
first resolved imaging of the radio photosphere of an S-type star to
allow comparison of its properties with M- and C-type stars and to
provide a benchmark for future studies.

\section{Sample Selection}
The targets for the current investigation include two M-type
(oxygen-rich) stars
previously observed by RM07 (R~Leo and W~Hya), the carbon star IRC+10216 
(previously imaged by M12), and the S-type\footnote{S-type stars
  contain similar amounts of carbon and oxygen.} star
$\chi$~Cyg. The centimeter-wavelength emission of
 $\chi$~Cyg  was studied at 8.4~GHz, 14.9~GHz, and 22.4~GHz
by RM97, but its radio photosphere has never been resolved. 
Table~1 summarizes some key
properties of the sample. R~Leo, $\chi$~Cyg, and IRC+10216 are all 
Mira-type (long-period) variables. W~Hya is
  often classified as a semi-regular variable, although it has just a
  single dominant pulsation period 
and its properties are generally
  similar to those of a Mira.\footnote{See
  {\url{https://www.aavso.org/vsot$\_$whya}} for 
discussion.}

%
\begin{deluxetable*}{llllccccl}
\tabletypesize{\tiny}
\tablewidth{0pc}
\tablenum{1}
\tablecaption{Summary of Target Stars and Observing Dates}
\tablehead{
\colhead{Name} & \colhead{$\alpha$(J2000.0)$^{*}$} &
\colhead{$\delta$(J2000.0)$^{*}$} 
 & \colhead{$V_{\star,\rm LSR}$} &  \colhead{$d$}  &
\colhead{Spectral} & \colhead{$P$} & \colhead{SiO} & \colhead{Date of}
  \\
\colhead{}     & \colhead{}     & \colhead{} & 
\colhead{(\kms)} & \colhead{(pc)}   & \colhead{Class} & \colhead{(days)}
& \colhead{Masers?} & \colhead{Observation}   \\
  \colhead{(1)} & \colhead{(2)} & \colhead{(3)} &
\colhead{(4)} & \colhead{(5)} & \colhead{(6)} & \colhead{(7)}
& \colhead{(8)} & \colhead{(9)}
}
\startdata
\tableline
\multicolumn{9}{c}{Oxygen-rich} \\
\tableline
R Leo & 09 47 33.4879 & 11 25 43.665 & $-1.0$ & 95 & M6e-9e&
303.5 & Yes&23-Feb-2014 \\
W Hya & 13 49 01.9981 & $-$28 22 03.488 & +42.0 & 110 & M7.5e-9e &
414.7 & Yes& 24-Feb-2014\\

\tableline
\multicolumn{9}{c}{S-type} \\
\tableline
$\chi$~Cyg & 19 50 33.9244 & 32 54 50.610 & +8.9 & 135 & S6.2e-10.4e& 409.3
&Yes & 13-Mar-2014\\

\tableline
\multicolumn{9}{c}{Carbon-rich}\\
\tableline
IRC+10216 & 09 47 57.4443(7) & 13 16 43.815(10)
& $-25.5$ & 130 & C9.5e & 630.0
&No & 22-Feb-2014
\enddata

\tablenotetext{*}{For R~Leo, W~Hya, and $\chi$~Cyg, the tabulated
  right ascension and declination are the nominal epoch J2000
  coordinates taken from the SIMBAD database (\url{http://simbad.harvard.edu}), uncorrected for
  proper motion. For IRC+10216, coordinates were measured from the
  observations (epoch 2014.145; see Appendix); the number in
  parenthesis is the uncertainty in the last significant digit. 
Absolute coordinates could not be
  measured for the other three stars because of the self-calibration
  procedures used to calibrate their phases (see Section~\ref{masercal}). }
\tablecomments{Units of right ascension are hours, minutes, and
seconds. Units of declination are degrees, arcminutes, and
arcseconds.
Explanation of columns: (1) star name; (2) \& (3) right
ascension and declination; (4) 
systemic velocity relative to the Local Standard of
Rest (LSR); (5) adopted distance in parsecs; (6) 
spectral type; (7) pulsation period in days; (8) indication of whether or not
the star has SiO maser emission; (9) date of observation. 
Adopted distances for R~Leo,
W~Hya, and $\chi$~Cyg are based on the period luminosity relation of
Feast et al. (1989) using the data of Haniff et al. (1995); the value
for IRC+10216 is based on the mean of literature values (see M12). Spectral types are from
Baudry et al. 1990 and Cohen 1979. The period of IRC+10216 is taken
  from M12; for the other three stars, periods were estimated based on fits to visible
  light curve data from the American Association of Variable Star
  Observers (AAVSO) using the {\sf VStar} software developed by 
D. Benn (\url{https://www.aavso.org/vstar}). Fits were performed
  using data from the three pulsation cycles bracketing the
  date of the VLA observations.}
\end{deluxetable*}


\section{Observations\protect\label{observations}}
Continuum observations of the four target stars (Table~1) were conducted 
at a central frequency of 46~GHz ($\lambda\sim7$~mm) using
the Karl G. Jansky Very
Large Array (VLA)\footnote{The VLA of the National Radio Astronomy
  Observatory (NRAO) is operated by Associated
  Universities, Inc. under cooperative agreement with the National
  Science Foundation.}  in its most extended (A) configuration (0.68~km
to 36.4~km
baselines). Observations of each star were obtained during a single
3-hour session. Data were recorded with 2-second integration times.
Antenna pointing corrections were evaluated hourly
using observations of a strong point source at X-band (8~GHz).  
3C286 was observed once per session to serve as a bandpass calibrator
and to allow absolute
calibration of the flux density scale. 

Two different observing strategies were employed, depending on whether or
not the target star has known SiO maser emission. 
For the stars with SiO masers (see Table~1), emission from the
strong SiO $v$=1,
$J$=1-0 line at 43.1~GHz was used to
calibrate the atmospheric phase variations, as was done previously 
by Reid \& Menten (1990) using H$_{2}$O masers and by RM97 using SiO masers. 
A 3-bit observing
mode with dual circular polarizations 
was used, and the WIDAR correlator was configured in a standard
``continuum'' mode, with 4 baseband pairs tuned to contiguously cover
an 8-GHz frequency window centered near 46~GHz. 
Each baseband pair contained 16
subbands, each with a bandwidth of 128~MHz and 128 spectral channels.
This frequency range included the SiO $v$=1 and $v$=2,
$J$=1-0 maser transitions as well as several other weaker SiO
lines. Each star was observed during two blocks of roughly
0.75 to 1 hour duration each, bracketed by observations of a
neighboring complex gain
calibrator (see Table~3). 

For IRC+10216, which is not an SiO maser emitter, the same 3-bit
correlator set-up was used as described above, with a center frequency
of $\sim$44~GHz. However, to allow calibration of the
atmospheric phases, rapid switching was performed performed between
the target and a nearby gain calibrator (see Table~2) with a duty cycle of
approximately 95 seconds (55 seconds on target, 40 seconds on a
calibrator).  The gain calibrator
scans  alternated between J1002+1216 
and J0943+1702 (lying 3.78~deg and 3.92~deg, respectively, from
IRC+10216). Weather conditions during our observations were clear and
dry, with wind speeds of $\sim$2.8~m s$^{-1}$. Carilli \& Holdaway (1997) have shown that this
combination of duty cycle, calibrator separation, and stable weather
conditions is generally sufficient to
result in near diffraction-limited seeing at 7~mm in the VLA A configuration
(see also below). 

\section{Data Reduction\protect\label{reduction}}
Data processing was performed using the Astronomical Image Processing
System (AIPS; Greisen 2003). 
The data were loaded directly into AIPS from archival
science data model 
files via the Obit software package (Cotton
2008). The default calibration (`CL') table was subsequently
regenerated  to  update the gain and
  opacity information, and antenna positions were updated to the best
  available values.

After 
flagging visibly corrupted data, a requantizer gain correction was
applied using the AIPS program {\small\sc{TYAPL}}. A fringe fit
  was then performed  using a 1-minute segment
    of data on 3C286 to correct the
  instrumental delays. Delay solutions were
  determined separately for the four independent basebands.
Bandpass calibration was performed in the standard manner, and the
absolute flux density scale was calculated by adopting 
the latest time-dependent flux density values for
3C286 from Perley \& Butler (2013), giving a flux
density as a function of frequency of the form: ${\rm log}(S_{\nu}) = 1.2515 - 0.4605({\rm
  log}(\nu)) - 0.1715({\rm log}(\nu))^{2} +0.0336({\rm log}(\nu))^3$, where
  $\nu$ is the frequency expressed in GHz and $S_{\nu}$ is in Jy. 

Following this step, different calibration procedures were
followed depending on whether or not observations of SiO maser emission were obtained.
These two cases are described in turn in the following subsections.

%
\begin{deluxetable*}{lrrccl}
\tabletypesize{\tiny}
\tablewidth{0pc}
\tablenum{2}
\tablecaption{Gain Calibration Sources}
\tablehead{
\colhead{Source} & \colhead{$\alpha$(J2000.0)} &
\colhead{$\delta$(J2000.0)} & \colhead{Flux Density (Jy)} &
\colhead{$\nu$ (GHz)} & \colhead{Date}
}

\startdata

J0935+0915$^{a}$ & 09 35 13.6411 & 09 15 07.813 & 0.128$\pm$0.003 &41.9395 &23 February 2014\\
      ...        &     ...       &  ...         & 0.117$\pm$0.003 &  49.8595   &   ...      \\ 

J0943+1702$^{b}$ & 09 43 17.2243 & 17 02 18.969 & 0.135$\pm$0.003 &40.0400 &22 February 2014 \\
...        & ...           & ...          & 0.115$\pm$0.003&47.9600 &...\\

J1002+1216$^{b}$ & 10 02 52.8452 & 12 16 14.587 & 0.100$\pm$0.003 &40.0400 & 22 February 2014  \\
...        & ...           & ...          & 0.096$\pm$0.002 &47.9600&...\\

J1339-2620$^{c}$ & 13 39 19.8907 & $-$26 20 30.496 & 0.379$\pm$0.006& 40.0395 & 24 February 2014\\
  ...               &    ...           &  ...               & 0.362$\pm$0.009& 47.9595 & ... \\  

J2010+3322$^{d}$ &20 10 49.7063 & 33 22 13.627&0.188$\pm$0.002 &41.9395& 13 March 2014\\
...              & ...          & ...         &0.154$\pm$0.002 &49.8595& ...
\enddata

\tablecomments{Units of right ascension are hours, minutes, and
seconds, and units of declination are degrees, arcminutes, and
arcseconds. Explanation of columns: (1) source name; (2) \& (3) right
ascension and declination (J2000.0); (4) derived flux density in Jy; values are quoted for the lowest and
highest frequency subbands from each observation;
(5) frequency at which the flux density in
the fourth column was computed; (6) date of observation.}

\tablenotetext{a}{Complex gain calibrator for R Leo.}
\tablenotetext{b}{Complex gain calibrator for IRC+10216.}
\tablenotetext{c}{Complex gain calibrator for W Hya.}
\tablenotetext{d}{Complex gain calibrator for $\chi$~Cyg.}

\end{deluxetable*}


\subsection{Calibration of Stars with Maser Emission\protect\label{masercal}}
For target stars where SiO maser emission was observed within the
band, the calibration approach of Reid \& Menten (1990; see also RM97) was adapted
for use with the new wide bandwidth VLA correlator (see Matthews
et al. 2015). In these cases, self-calibration on the bright maser
emission allows improvement of the calibration of the atmospheric phases,
effectively allowing achievement of nearly perfect ``seeing'' in the stellar
continuum measurements.

Following bandpass calibration and calculation of the absolute
flux density scale as described above, calibration of the
frequency-independent portion of the complex gains was performed for
the observed calibrator sources following the standard approach for
high-frequency data and their flux
densities were computed (see
Table~3). Amplitude and phase corrections computed from the phase
calibration source(s) were then applied to the data from the
target star, allowing the removal of any slow (hour timescale), instrumental gain
drifts. Additionally, corrections to the positions of the stars to
account for their proper motions were applied based on the values from van Leeuwen
(2007).  This ensured that the stellar
emission was located close to the phase center of subsequent images.

After these initial calibration steps, the spectral channel 
containing the strongest SiO $v$=1, $J$=1-0 maser emission from the
target star was split 
from the main data set, and several iterations of phase-only
self-calibration were performed until convergence was reached. Based on
these solutions, phase corrections appropriate for each of the 64
subbands across the full 8-GHz continuum band were derived using 
the AIPS task {\small\sc{SNP2D}}. 

Following application of the above corrections, a second round of
self-calibration was performed on the reference channel,
solving for both amplitudes and phases. To prevent
drift in the amplitude scale, the gains were normalized during this step. After these
corrections were applied to the full data set, the data were
averaged in time to 10-second records and 
spectral channels 1-3 and 125-128 at the edges of each
subband were flagged because of their significantly higher noise levels. Spectral
channels known to contain line emission or radio frequency interference 
were also flagged.  Next, the data were
Hanning-smoothed in frequency to suppress Gibbs
ringing artifacts caused by the narrow SiO maser lines.
At this stage, the AIPS task {\small\sc{REWAY}} was used to compute 
optimized weights for the visibility data, and lastly, the data were
further averaged in frequency to produce 8 spectral channels per subband.

\subsection{Calibration of Data without Maser Emission\protect\label{nomasercal}}
For the IRC+10216 data set, calibration of the
frequency-independent portion of the complex gains was performed in the
standard manner using
observations of the calibrators J0943+1702 and J1002+1216. First,
phase-only corrections were solved for and applied, followed by
amplitude and phase corrections. The typical residual rms
scatter in the phase solutions was $\lsim$6~deg for all antennas.

Subsequent steps were the same as for
the sources with SiO lines, with the exception that flagging of line
emission was unnecessary, as no
significant line emission
was found within the observing band (39.98-48.02~GHz). Prior to imaging
(Section~\ref{imaging}), a positional offset was applied to
compensate for proper motion and place the star at the phase center. 

The self-calibration procedure used for the stars with SiO maser
emission 
(Section~\ref{masercal}) destroys absolute position
information.  However, the calibration technique used for IRC+10216
preserves the positional accuracy and allowed us to measure the absolute
position of the star and improve the characterization of its proper
motion. This analysis is described in the Appendix.

\subsection{Imaging the Data\protect\label{imaging}}
Initial imaging and deconvolution of the visibility data were performed using
the {\small\sc{CLEAN}} deconvolution algorithm as implemented in the 
AIPS {\small\sc{IMAGR}} program (but see Section~\ref{sparse}). For
the {\small\sc{CLEAN}} images presented
here, we used a Briggs robustness parameter of ${\cal R}$=0, a cell size of
5~mas, and a circular restoring beam
(see Table~3). Corrections were applied during imaging for the
frequency dependence of the primary beam and the expected spectral
index of the sources ($\alpha$=1.86; RM97). Imaging was also attempted
using a multi-scale {\small\sc{CLEAN}} algorithm, but differences in the resulting
images were
insignificant.


%
\begin{deluxetable}{lccrcc}
\tabletypesize{\tiny}
\tablewidth{0pc}
\tablenum{3}
\tablecaption{Restoring Beam Parameters}
\tablehead{
\colhead{Star} & 
\colhead{$\theta_{\rm a}$} & \colhead{$\theta_{\rm b}$} & \colhead{PA} &
\colhead{$\theta_{\rm circ}$} & \colhead{RMS noise}\\
\colhead{} & \colhead{(mas)} & \colhead{(mas)} &
\colhead{(degrees)}& \colhead{(mas)} & \colhead{($\mu$Jy beam$^{-1}$)}
}

\startdata
R Leo & 42 & 34 & $-27.8$ & 38 & 16.4\\
W Hya & 82 & 34 & $-9.7$ & 70 & 21.1\\
$\chi$ Cyg & 42 & 37 & $-72.0$ & 39& 13.6\\
IRC+10216 & 41 & 36 & $-23.6$ &38 & 15.7 

\enddata

\tablecomments{Images summarized here were produced using the {\sc{CLEAN}} deconvolution
  method with robust weighting (${\cal R}$=0) (see
  Section~\ref{imaging}). 
$\theta_{\rm a}$ and $\theta_{\rm b}$ are the
major and minor axes, respectively, of the dirty beam, measured at
  FWHM. The position angle (PA) of the beam was measured east from
  north.
The images used for the present analysis
were produced using circular restoring beams with FWHM diameters
  $\theta_{\rm circ}$ equal to the geometric mean
  of the dirty beam, except for W~Hya, where a circular beam with FWHM
  70~mas was adopted because of the significant beam elongation.}

\end{deluxetable}


%
\begin{deluxetable*}{lllllllll}
\tabletypesize{\tiny}
\tablewidth{0pc}
\tablenum{4}
\tablecaption{Measured Stellar Parameters at 46 GHz (7~mm)}
\tablehead{
\colhead{Star} &  \colhead{Optical Phase} & \colhead{$\theta_{\rm
    maj}$} & \colhead{$\theta_{\rm min}$} &
\colhead{PA} & \colhead{$e$} & \colhead{$S_{\nu}$} &
\colhead{$D$} & \colhead{$T_{b}$}\\
\colhead{} & \colhead{} &  \colhead{(mas)} & \colhead{(mas)} &
\colhead{(degrees)} & \colhead{} 
& \colhead{(mJy)}    & \colhead{(AU)} & \colhead{(K)}\\
  \colhead{(1)} & \colhead{(2)} & \colhead{(3)} & \colhead{(4)} & \colhead{(5)}
& \colhead{(6)} & \colhead{(7)} & \colhead{(8)} & \colhead{(9)}  }

\startdata
\tableline
\\
\multicolumn{8}{c}{Uniform Elliptical Disk Fits to
  Visibilities (New Data)}\\
\\
\tableline

R Leo & 0.23 & 54$\pm$3 (0.4)& 45$\pm$2 (0.5) & 109$\pm$5
(1.1) & 0.17$\pm$0.06 & 5.8$\pm$0.9 (0.02) & 4.7$\pm$0.2 & 2016$\pm$340\\

W Hya & 0.61 & 71$\pm$4 (1.0) & 65$\pm$3 (0.4) & 0$\pm$5
(1.8) & 0.09$\pm$0.07 & 9.0$\pm$1.4 (0.03) & 7.5$\pm$0.3 & 1645$\pm$280\\

$\chi$~Cyg & 0.75 &39$\pm$2 (1.3) & 36$\pm$2 (1.1) &
133$\pm$11 (9.5) & 0.07$\pm0.08$ & 2.2$\pm$0.3 (0.02) & 5.1$\pm$0.2 & 1327$\pm$210  \\

IRC+10216 & 0.41$^{a}$ & 88$\pm$4 (1.2) & 75$\pm$4 (1.1)
&70$\pm$5 (0.43) & 0.15$\pm$0.06 & 11.4$\pm$1.7 (0.01) & 10.6$\pm$0.4 & 1580$\pm$260
\\
\tableline
\\
\multicolumn{8}{c}{Elliptical Gaussian Fits to Images
  (New Data)}\\
\\

\tableline

R Leo & 0.23 & 34$\pm$2 (0.2) & 29$\pm$1 (0.2) &
101$\pm$6 (3) & 0.15$\pm$0.06 & 6.0$\pm$0.9 (0.04) & ...& ...\\

W Hya & 0.61 & 42$\pm$2 (0.2) & 39$\pm$2 (0.2) & 14$\pm$9
(8) & 0.07$\pm$0.06 & 9.2$\pm$1.4 (0.04) & ...& ... \\

$\chi$~Cyg & 0.75 & 22$\pm$1 (0.4) & 22$\pm$1 (0.4) &
171$\pm$56 (56) & 0.0$\pm$0.09 & 2.2$\pm$0.3 (0.03) & ...& ... \\

IRC+10216 & 0.41$^{a}$ & 61$\pm$4 (0.3) & 52$\pm$4 (0.2)
& 74$\pm$5 (1) & 0.15$\pm0.06$ & 12.0$\pm$1.8 (0.06) & ... &
... \\

\tableline
\\
\multicolumn{8}{c}{Uniform Elliptical Disk Fits to
  Visibilities (Previous Epochs$^{b}$)}\\
\\
\tableline
R Leo & 0.55 & 61$\pm$10 & 39$\pm$6 & $160\pm$12& 0.36$\pm$0.14 &
4.1$\pm$0.2 & 4.6$\pm$0.5 & 1630$\pm$410\\

W Hya & 0.25 & 69$\pm$10 & 46$\pm$7 & 83$\pm$18 &0.33$\pm$0.14 &
8.0$\pm$0.4 & 6.2$\pm$0.6 & 2380$\pm$550\\

$\chi$~Cyg & 0.09 & ... & ... & ... & ... & $\sim$4.5$^{c}$ & ... & ...\\

IRC+10216 & 0.79$^{a}$ & 87$\pm$2 & 80$\pm$ 1 & 22$\pm$5 & 0.08$\pm$0.02 &
12.2$\pm$0.1 & 10.9$\pm$0.1 & 1660

\enddata

\tablenotetext{a}{For IRC+10216, an IR phase is quoted; for
  M-type Miras, the IR phase typically lags the optical phase by $\approx$0.1--0.2,
  but the offset is not well-established for carbon stars (Smith et
  al. 2006; M12).}
\tablenotetext{b}{Previous measurements are taken
  from RM07 (R~Leo, W~Hya, and $\chi$~Cyg) and M12 (IRC+10216); quoted uncertainties
 are taken from the original references and do not include
calibration  uncertainties in the absolute flux density scale or the
  full range of systematic uncertainties included in the present study
  (see Matthews et al. 2015 for discussion).}
\tablenotetext{c}{Estimated 46~GHz flux density based on extrapolation
  from unresolved measurements
  at 8.44~GHz, 14.9~GHz, and 22.4~GHz.}
\tablecomments{Pulsation phases for R~Leo, W~Hya, and $\chi$~Cyg were
  computed by fitting visual light curve data from the AAVSO database
(see Notes to Table~1). For IRC+10216, the phase was estimated 
from the infrared data of Shenavrin, Taranova, \&
  Nadzhip (2011) through an extrapolation based on
  the period and date of maximum light derived by M12 (630~days and JD$_{\rm max}$ =
  2452554, respectively). Quoted
  uncertainties on all quantities include formal, systematic, and calibration errors
  (see Text for details). Uncertainties in derived quantities do not
  account for uncertainties in the distance. For measured quantities, the value in
  parentheses indicates the contribution to the error budget from
  formal fit uncertainties.  For a uniform disk,
$\theta_{\rm maj}$ and $\theta_{\min}$ are the major and minor axis sizes of
the disk, respectively; for a Gaussian fit they represent the
the FWHM dimensions of the elliptical Gaussian, after deconvolution with the size
of the dirty
beam. For a resolved source, $\theta_{\rm maj}$ measured from a  Gaussian fit is
expected to be
$\approx$0.625 times the value measured from a uniform
  disk fit.  Explanation of columns: (1) star name; (2) optical
  pulsation phase; (3) FWHM diameter of the major axis i mas; (4) FWHM
  diameter of the minor axis in mas; (5) position angle of the major axis in
  degrees, measured east from north; (6) ellipticity, defined as 
  $e=(\theta_{\rm maj} - \theta_{\rm min})/(\theta_{\rm maj})$; (7) 46~GHz
  flux density in mJy; (8) mean diameter of the radio
  photosphere in AU, derived using the geometric mean angular
  diameter; (9) brightness
  temperature, which for a uniform disk is defined as $T_{b}=2S_{\nu}c^2/(k\nu^{2}\pi\theta_{\rm
    maj}\theta_{\rm min})$ where $c$ is the speed of light, $k$ is
  the Boltzmann constant, and all quantities are expressed in cgs
  units. }

\end{deluxetable*}



%
\begin{deluxetable*}{lclccl}
\tabletypesize{\tiny}
\tablewidth{0pc}
\tablenum{5}
\tablecaption{Temporal Changes in Stellar Parameters}
\tablehead{
\colhead{Star} & \colhead{Elapsed Time} & \colhead{$\Delta\theta_{\rm
    maj}$} & \colhead{$\Delta\theta_{\rm min}$} &
\colhead{$\Delta$PA} & \colhead{$\Delta S_{\nu}$}\\
\colhead{} & \colhead{(days)} & \colhead{(mas)} & \colhead{(mas)} &
\colhead{(degrees)}
& \colhead{(mJy)}    \\
  \colhead{(1)} & \colhead{(2)} & \colhead{(3)} & \colhead{(4)} & \colhead{(5)} &\colhead{(6)} }

\startdata

R Leo & 4868 & $-7\pm10$ & $+6\pm6$ & ${\bf -51\pm13}$ & ${\bf +1.7\pm0.9}$\\

W Hya & 4862 & $+2\pm11$ & ${\bf +19\pm8}$ & ${\bf -83\pm19}$ & $+1.0\pm1.5$\\

$\chi$ Cyg & 8811 & ... & ... & ... & ${\bf -2.3\pm0.9}$ \\

IRC+10216 & 2918 & $+1\pm4$ & ${\bf -5\pm4}$ & ${\bf +48\pm7}$ & $-0.8\pm1.7$

\enddata

\tablecomments{Observed changes in stellar parameters between
  the current observing epoch (this paper) and earlier
  epochs, when available. Values are based on the fits to the visibility data summarized
  in Table~4. Parameters with statistically significant changes are
  highlighted in boldface. For $\chi$~Cyg, a 20\% uncertainty in 
the flux density was assumed for the earlier epoch.}

\end{deluxetable*}

%
\begin{figure}
\centering
\scalebox{0.41}{\rotatebox{0}{\includegraphics{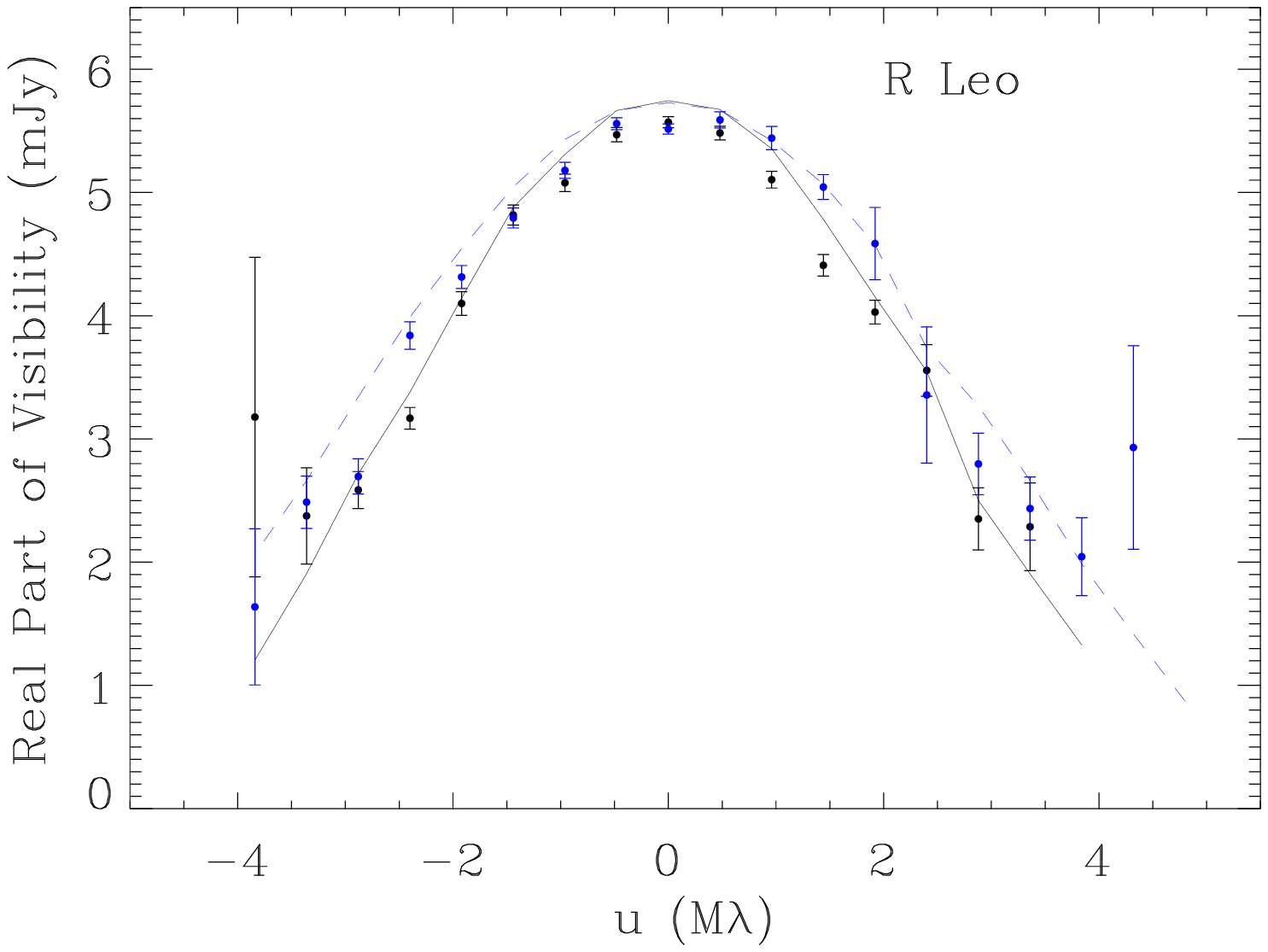}}}
\scalebox{0.41}{\rotatebox{0}{\includegraphics{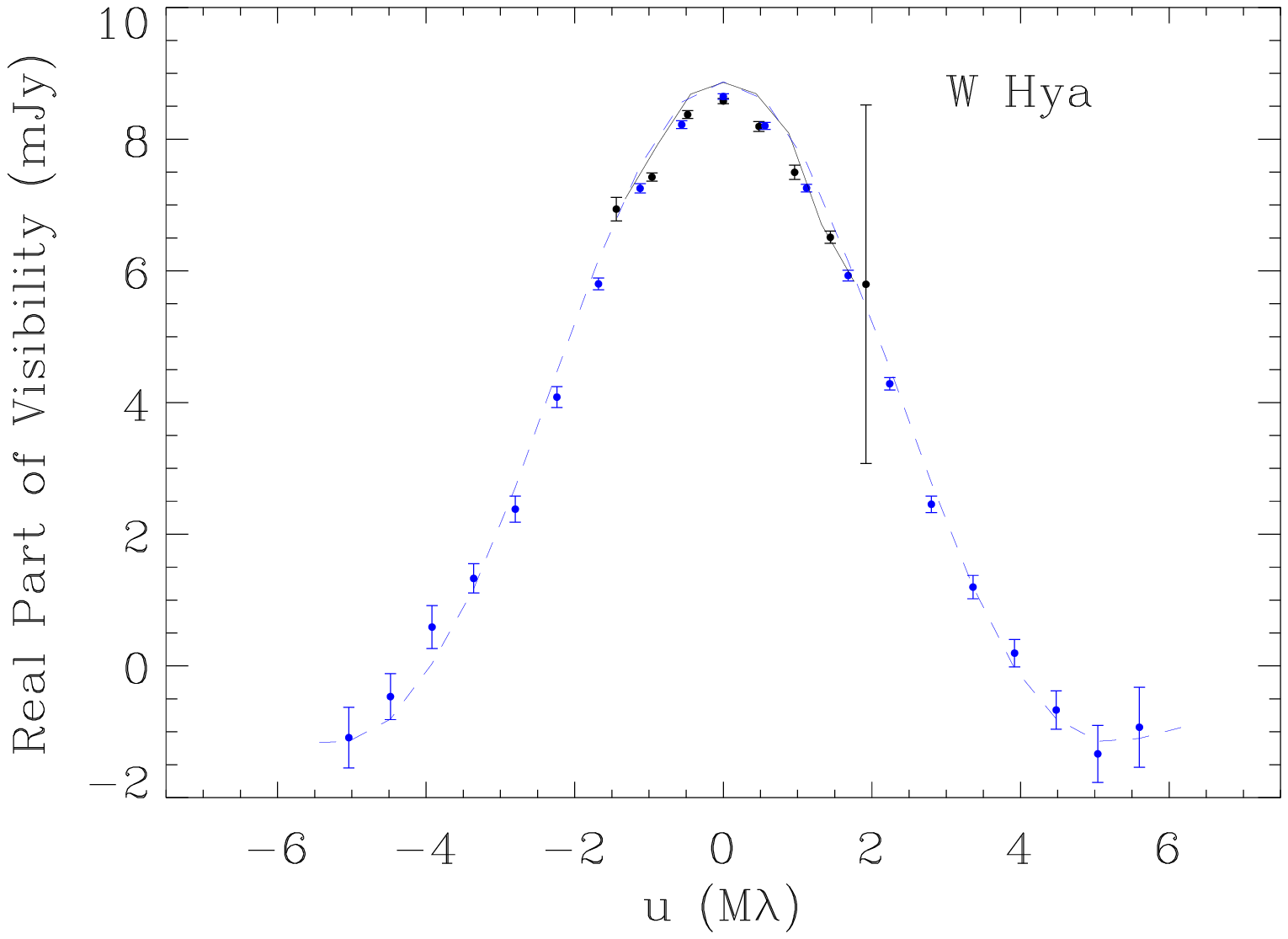}}}
\scalebox{0.41}{\rotatebox{0}{\includegraphics{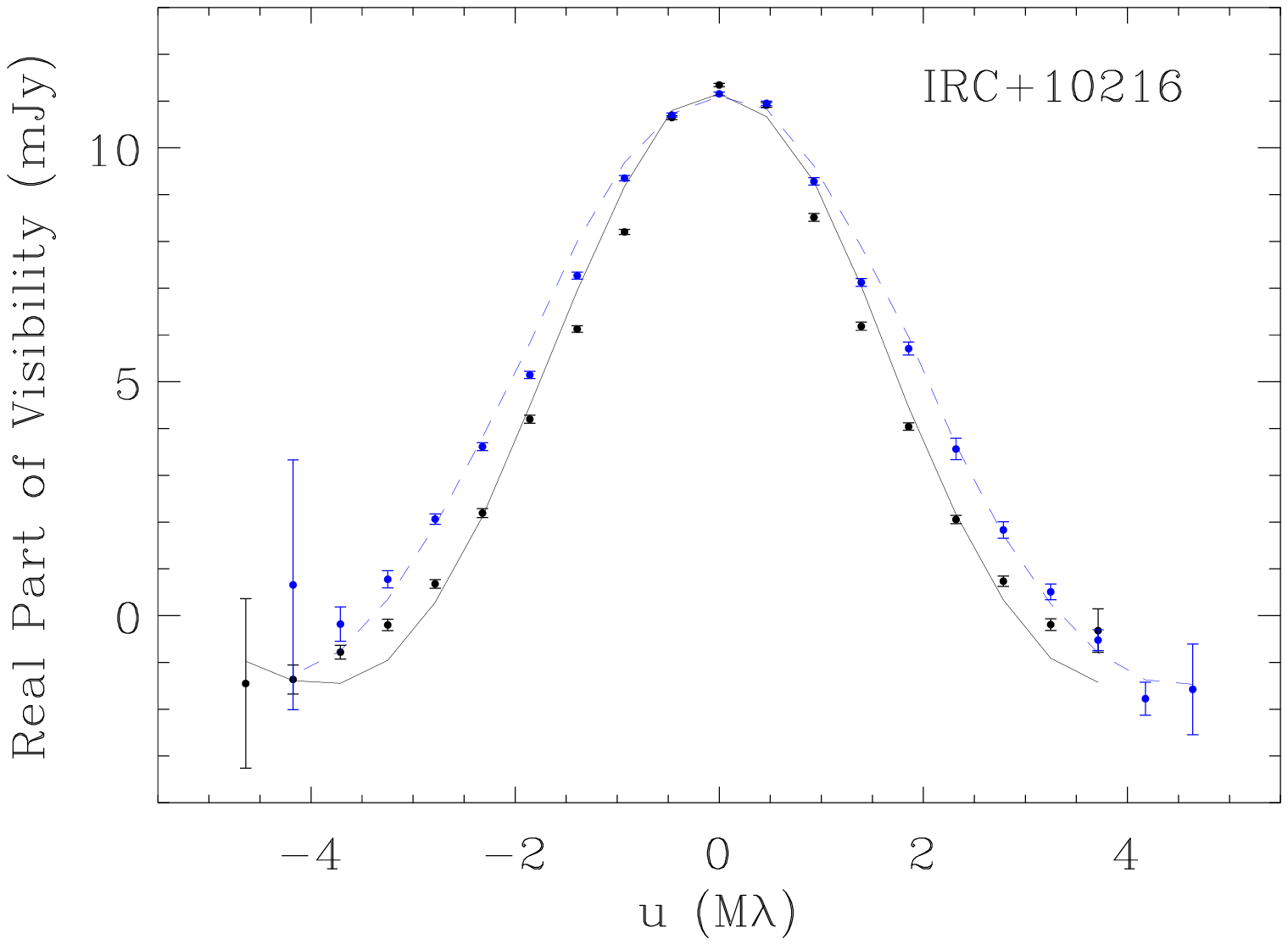}}}
\scalebox{0.41}{\rotatebox{0}{\includegraphics{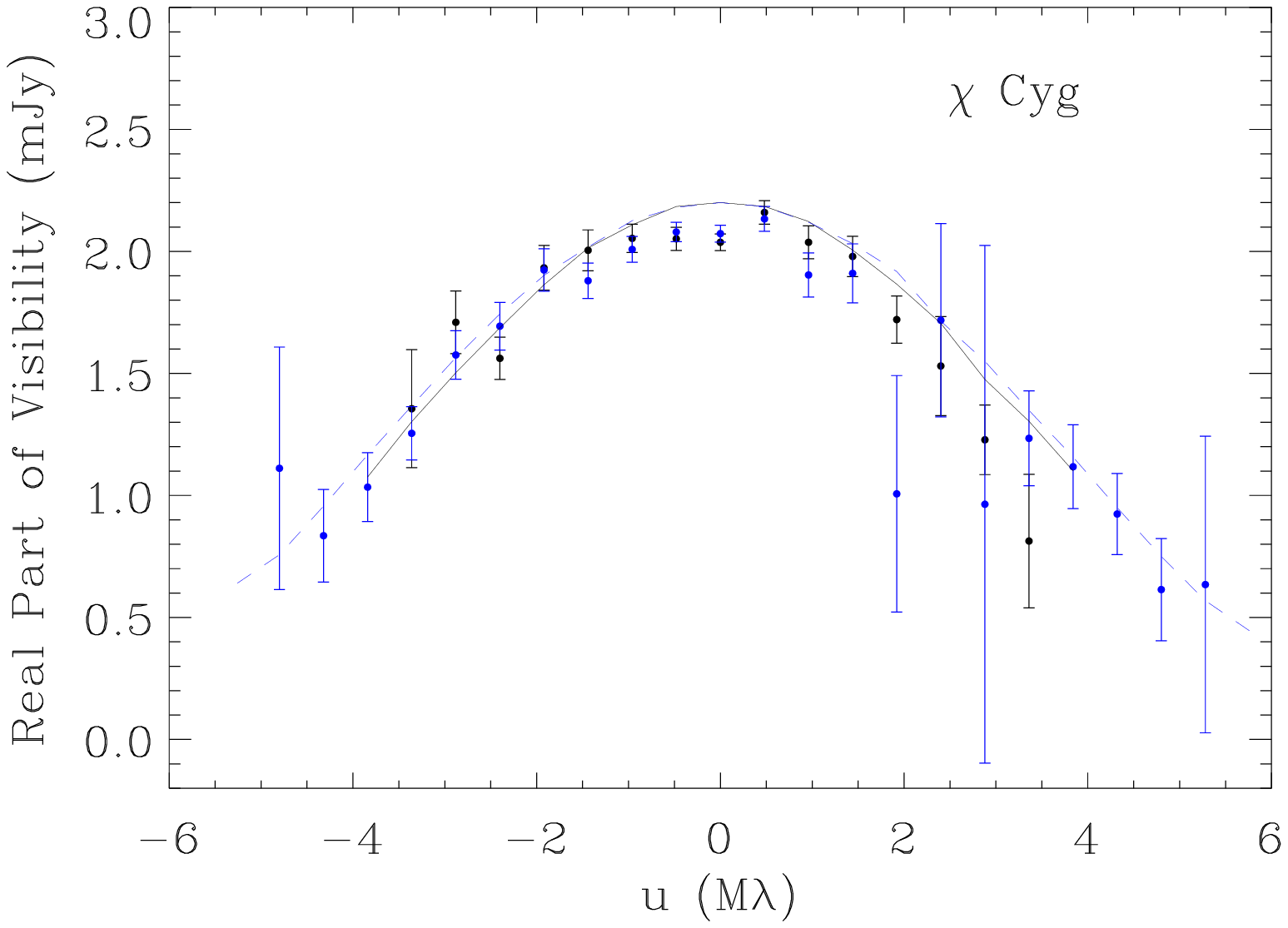}}}
\caption{Real part of the visibility amplitude along the major axis
  (black points) and minor axis (blue points) of the stellar disks.  
In each case, the visibility data were rotated
  using the PA values from Table~4
  to align the major and minor disk axes, respectively, with the
  rotated $u$-axis
  of the $u$-$v$ plane. 
Brightness cuts were then extracted along the $u$-axis
  using binned averages of data from a
  narrow strip along the $v$-axis ($|v|<0.5$M$\lambda$). 
  The black
  solid (blue dashed) lines show the major (minor) axis
  profiles based on the best-fitting 2D uniform
  elliptical disk model parameters from Table~4 (they are not fits to
  the plotted points). }  
\label{fig:uvcuts}
\end{figure}

\section{Results}
\subsection{Measured 
Radio Photosphere Parameters for the Target Stars\protect\label{measurements}} 
\subsubsection{Evidence for
Deviations from Circular Symmetry\protect\label{visfits}}
To measure the size, shape, and flux densities  of the radio
photospheres of the four stars in our sample, 
we have fitted two-dimensional (2D) uniform elliptical disk models to
the visibility data using the AIPS task {\small\sc{OMFIT}}. The
results are presented in Table~4. 
The quoted uncertainties include contributions from  the formal fitting uncertainties
(Condon 1997), as well as from calibration and
systematic errors  (see
Appendix of Matthews et al. 2015 for details). The dominant source of uncertainty in the
derived flux densities is the  absolute
calibration uncertainty, which we assume to be 15\%
at 7~mm.\footnote{\url{https://science.nrao.edu/facilities/vla/docs/manuals/oss/performance/fdscale.}}
As a consistency check, we also determined the
stellar parameters of the sample based on elliptical Gaussian fits to the {\small\sc{CLEAN}} images
(Section~\ref{imaging}) using the AIPS task {\small\sc
  {JMFIT}}. In all cases the parameters derived from the two
fitting methods agree to 
within uncertainties. (For resolved sources, the FWHM sizes derived from Gaussian fits
are expected to be a factor of $\approx$0.625 times smaller than those from a
uniform disk fit). 

Based on the fit results in Table~4, the shapes of three of the stars in
the current sample 
show deviations from circular symmetry. 
To further illustrate this effect, in Figure~\ref{fig:uvcuts} we plot radial brightness cuts
along the major and minor axis of each  star in the visibility domain. To produce these profiles,
the visibility data were rotated to place the major and minor axis, respectively, of
each stellar disk along the rotated $u$-axis of the visibility
plane. The data used to form each plotted point along the $u$-direction 
were then constrained to a narrow
strip along the $v$-axis ($|v|<0.5$M$\lambda$). 

Both R~Leo and IRC+10216 exhibit statistically
significant differences between 
their major and minor axis brightness profiles, indicative of a non-zero
ellipticity.  On the other hand,
$\chi$~Cyg shows no statistically significant deviation from circular
symmetry. These results are all consistent with the
2D uniform
elliptical disk fits presented in Table~4.
In the case of
W~Hya, the ability to compare the major and minor axis brightness
profiles in Figure~\ref{fig:uvcuts} is limited by the relatively low
spatial
resolution along the major axis of
the dirty beam (see Table~3). 

\subsubsection{Evidence for
Secular Shape Changes\protect\label{secular}}
\begin{figure}
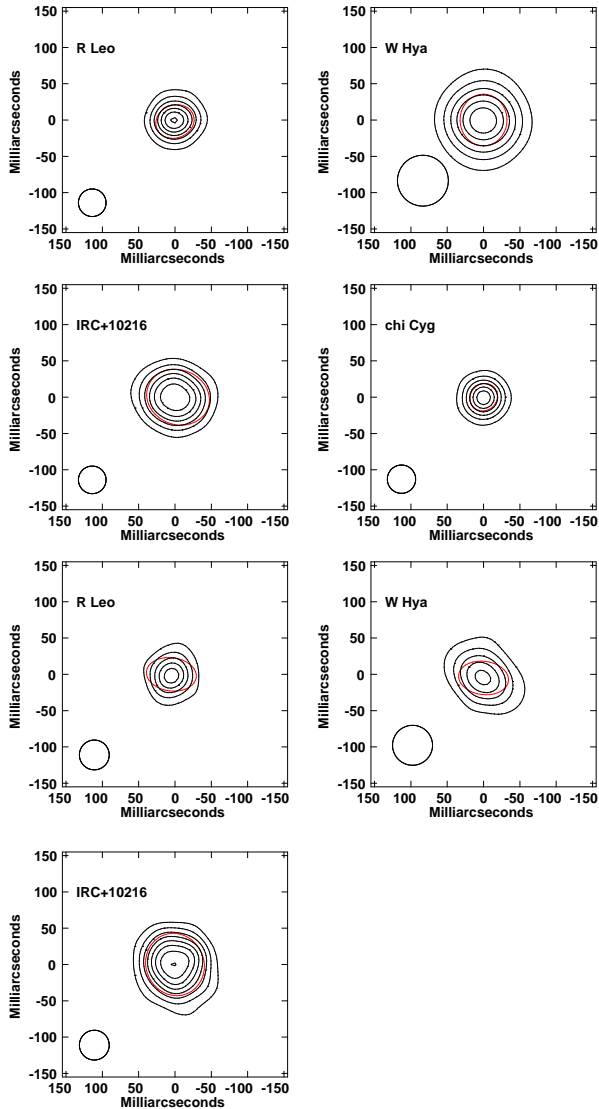

\centering
\scalebox{0.21}{\rotatebox{0}{\includegraphics{f2a.ps}}}
\scalebox{0.21}{\rotatebox{0}{\includegraphics{f2b.ps}}}
\scalebox{0.21}{\rotatebox{0}{\includegraphics{f2c.ps}}}
\scalebox{0.21}{\rotatebox{0}{\includegraphics{f2d.ps}}}
\scalebox{0.21}{\rotatebox{0}{\includegraphics{f2e.ps}}}
\scalebox{0.21}{\rotatebox{0}{\includegraphics{f2f.ps}}}
\scalebox{0.21}{\rotatebox{0}{\includegraphics{f2g.ps}}}
\scalebox{0.21}{\rotatebox{0}{\includegraphics{f2h.ps}}}
\caption{{\it Top row:} {\small\sc{CLEAN}} contour images of the four stars observed with the
 VLA in 2014: R~Leo, W~Hya, IRC+10216, and $\chi$~Cyg. The restoring
 beam is indicated in the lower left corner. The red ellipse indicates
 the best-fitting uniform elliptical disk model based on fits to the
 visibility data (Section~\ref{visfits}). Contour levels are
 (1,2,3,...8)$\times$0.5~mJy beam$^{-1}$ for R~Leo and IRC+10216,
 (1,2,3...8)$\times$1.0~mJy beam$^{-1}$ for W~Hya,
 and (1,2,3...8)$\times$0.25~mJy beam$^{-1}$ for
 $\chi$~Cyg. {\it Bottom row:} Contour
 images of the same stars based on data from previous epochs, when
 available (see
 Table~4 and Section~\ref{visfits}). The contour levels for each star are
 identical to those in
 the upper panels.  }  
\label{fig:kntrmaps}
\end{figure}

In Figure~\ref{fig:kntrmaps} we present contour images of each of
the four observed stars (top row), with the results of the uniform
elliptical disk fits to the visibility data (Table~4) overplotted as red
ellipses. 
Along the bottom row of this figure we show comparable plots for the
three previously observed stars based on the data from RM07 and 
M12. For each individual star, the data from the two epochs are contoured
identically. 

An examination of Figure~\ref{fig:kntrmaps}, as well as the fit 
results in Tables~4 and 5,
show that there appear to be statistically significant changes in the size,
shape, and/or orientation of the radio photospheres of all three of
the stars that have been imaged previously (R~Leo, W~Hya, and
IRC+10216). 
In addition, both R~Leo
and $\chi$~Cyg appear to have undergone 
changes in flux density compared with 
earlier measurements.  While $\chi$~Cyg was not previously
observed at 7~mm, an extrapolation of its flux density from longer
wavelength measurements 
taken in January 1990, assuming a spectral index of 1.86 (RM97), predicts a
7~mm flux density a factor of two larger than we measure from our
recent observations.
In the subsection that follows, 
we present additional discussion of the results for each individual
star.

%

\subsection{Results for Individual Stars}
\subsubsection{R Leo}
R~Leo is one of two stars for which RM07 measured a significant
deviation from sphericity, with an ellipticity of 0.64$\pm$0.14.  Our
new radio observations find a somewhat rounder, though still elongated shape,
with a different orientation compared with the earlier data
(Table~5).   An ellipticity
$\sim$0.11 has also been reported previously for the optical
photosphere of R~Leo at variability phase $\phi$=0.71 by Lattanzi et
al. (1997), and a comparison with the earlier measurements by Tuthill et
al. (1994) suggests that the elongation of the optical photosphere also changes orientation
over time.  

Our new measurement of the 46-GHz flux density of R~Leo is $\sim$50\% higher
than the previous measurement of RM07. Based on 8.4-GHz measurements, RM97 found that radio
photosphere fluxes tend to vary by $\lsim$15\% over the course of a
pulsation cycle. However, at 46~GHz, flux density changes of order
50\% over multi-year 
timescales have also
been seen in measurements of Mira (Matthews \& Karovska 2006; Matthews et
al. 2015), suggesting that either the stars are more variable at
higher frequencies, or that intra-cycle changes of larger amplitude are
occurring, possibly due to changes in radio opacity (e.g., O'Gorman et
al. 2015).

Despite the significant change in the radio flux density of R~Leo between epochs,
the observed shape change of the star, and the fact that
the observations were conducted at different pulsation phases, the inferred mean diameter of
R~Leo is comparable during the two epochs. In contrast, at optical wavelengths,
Burns et al. (1998) found that the size of R~Leo changes by as much as
50\% over the course of the stellar pulsation cycle. At these shorter
wavelengths, the large changes are most likely primarily due to
opacity changes and/or
changes in the temperature structure of the outer atmosphere than
physical motions of the surface. 

Several authors have also previously published measurements of the size of R~Leo
based on observations in the near-infrared. For example, at 2.16$\mu$m,
Perrin et al. (1999) found a diameter of 
28.18$\pm$0.05~mas at a phase
$\phi$=0.24, consistent with Wittkowski et al. (2016), who
measured 29.6$\pm$1.3~mas at $\phi$=0.6.  
Several other near-infrared diameter measurements  in the literature are in general
agreement with these values (e.g., Mennesson et al. 2002; Monnier et
al. 2004;
Fedele et al. 2005). A comparison with these various measurements suggests
that the radio photosphere of
R~Leo is $\sim$1.7 times the extent of photosphere as measured at 2~$\mu$m.
Using mid-infrared data, Paladini
et al. (2017) also recently reported evidence of a non-zero differential phase
in R~Leo that they attribute to variable asymmetries.


%
\begin{figure*}
\centering
\scalebox{0.4}{\rotatebox{0}{\includegraphics{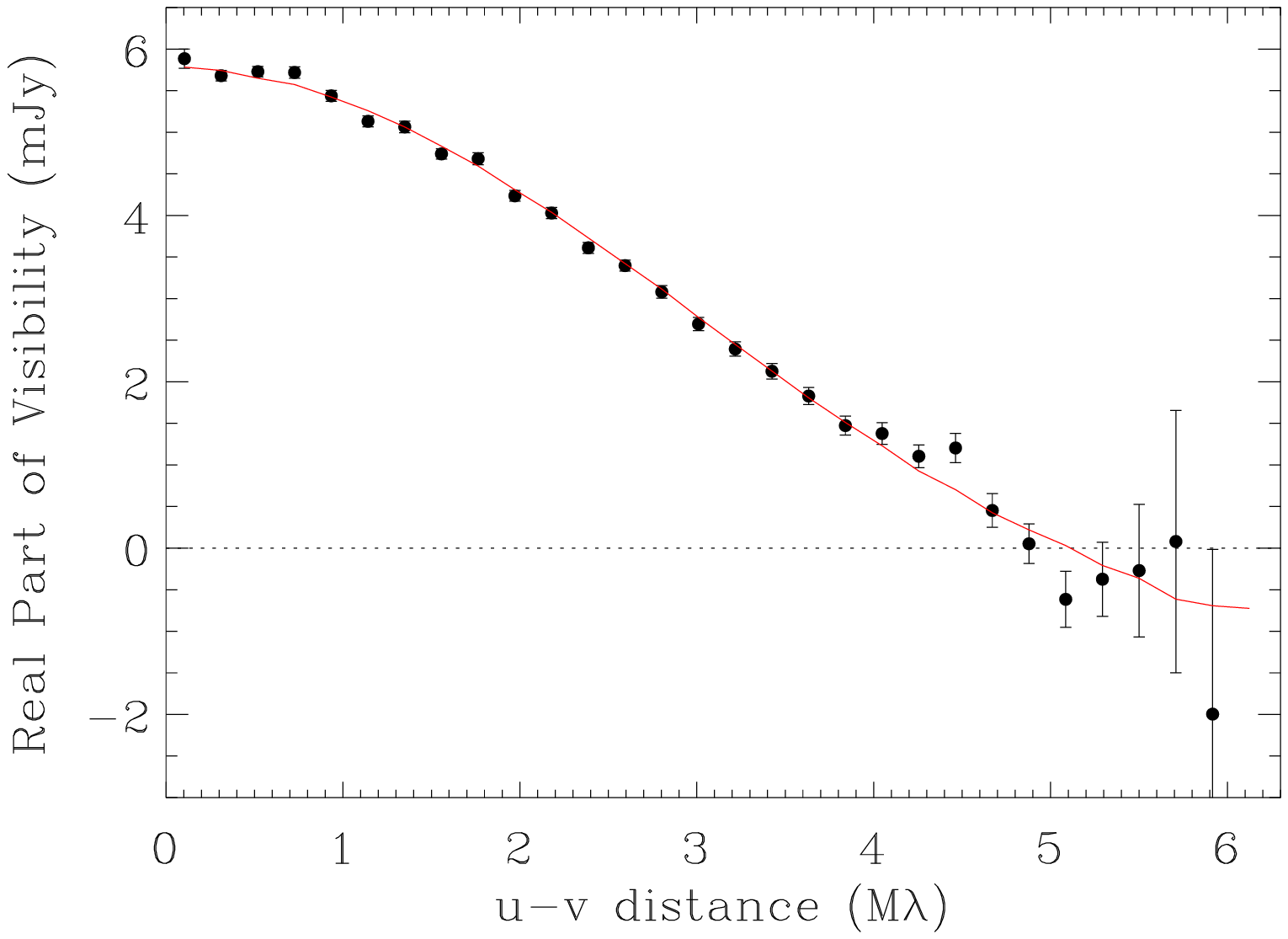}}}
\scalebox{0.4}{\rotatebox{0}{\includegraphics{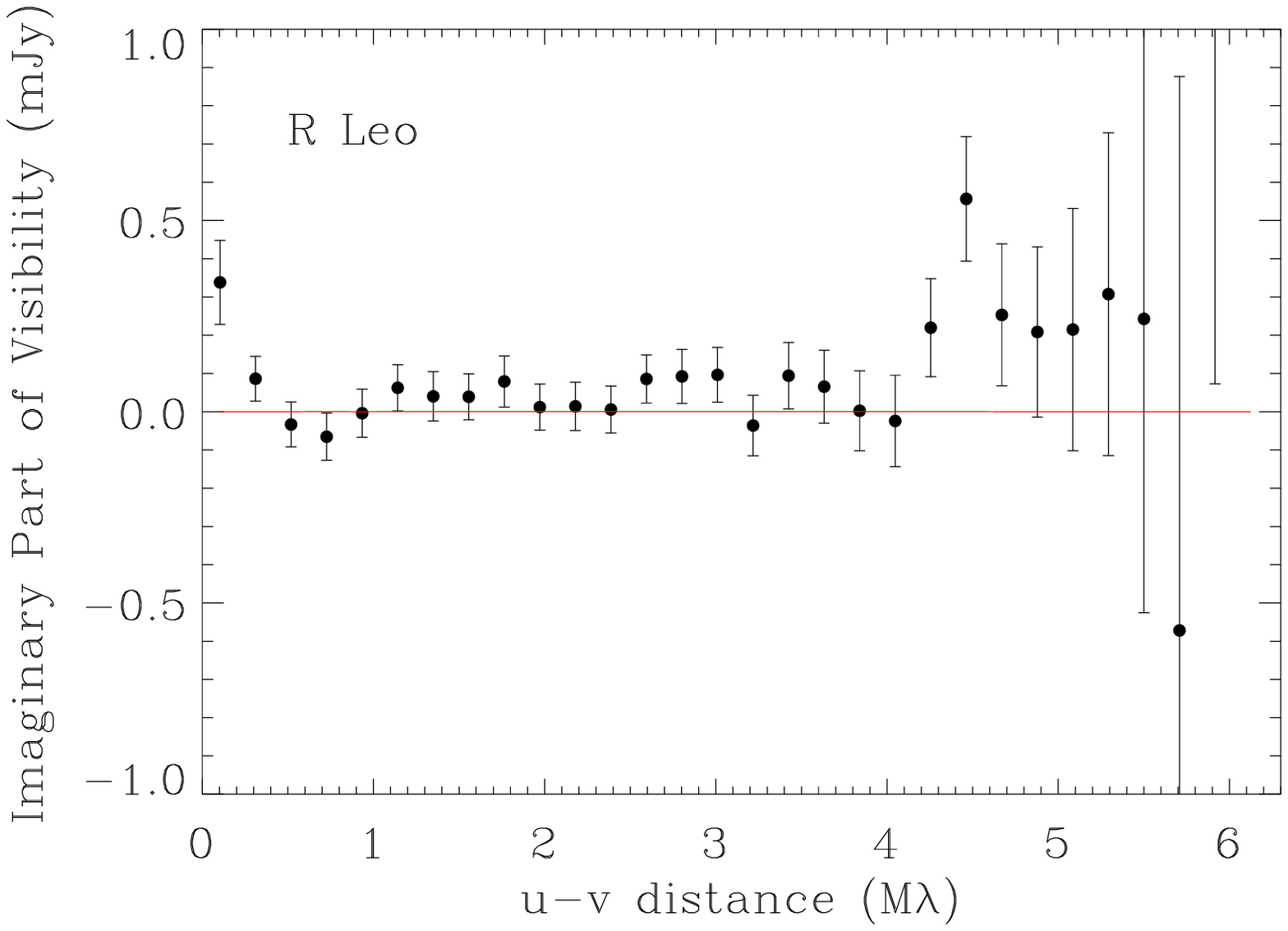}}}
\scalebox{0.4}{\rotatebox{0}{\includegraphics{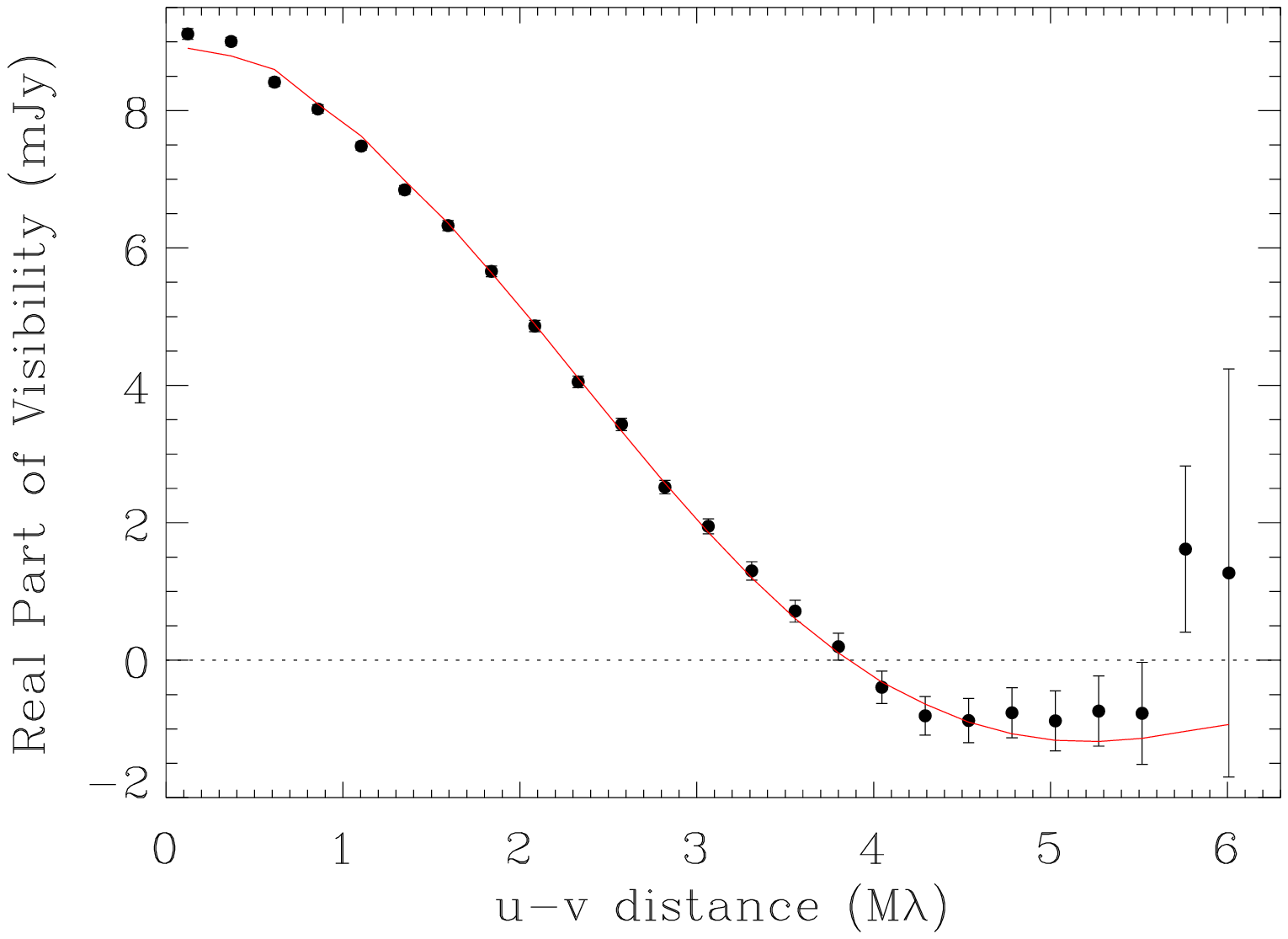}}}
\scalebox{0.4}{\rotatebox{0}{\includegraphics{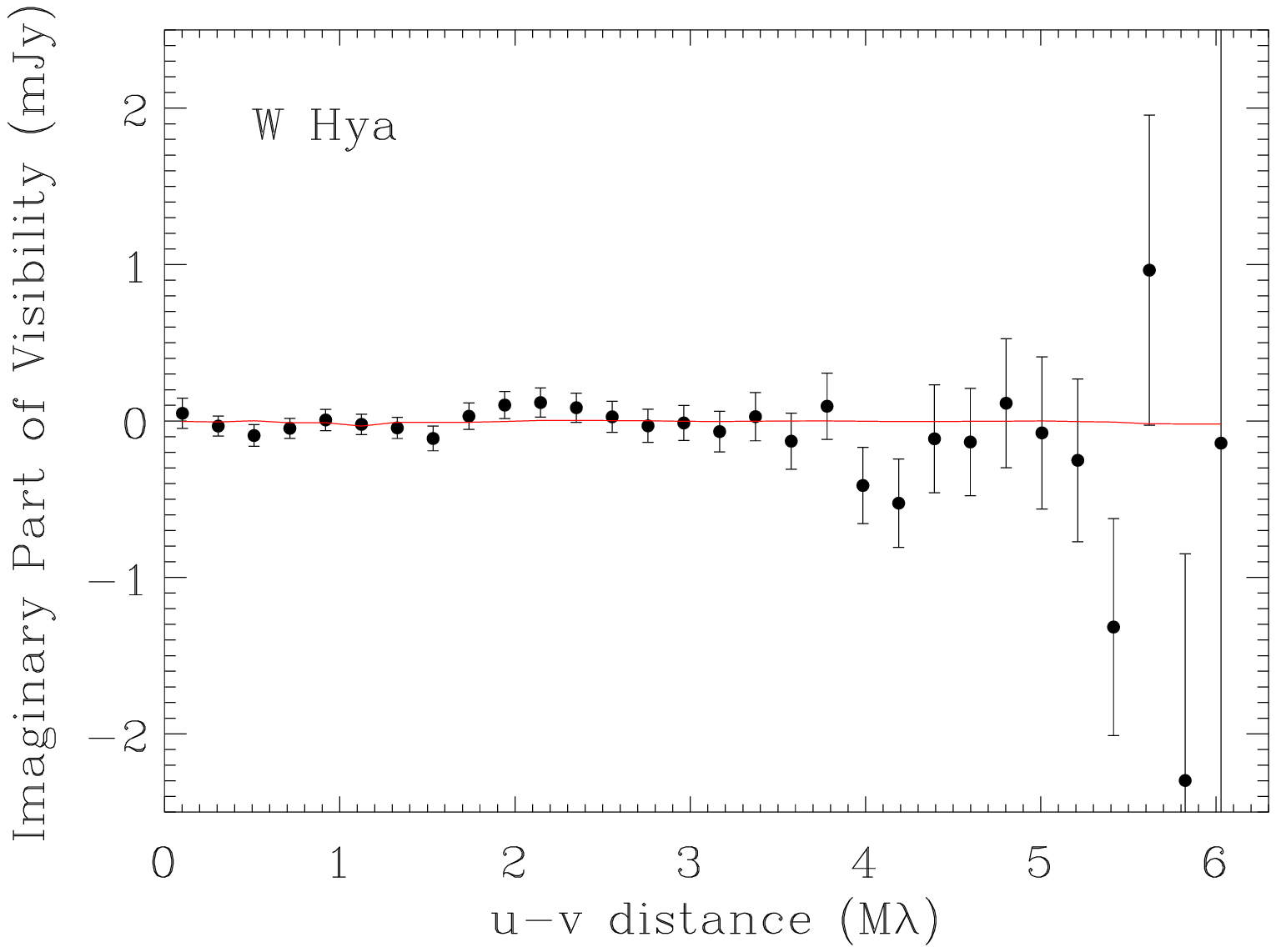}}}
\scalebox{0.4}{\rotatebox{0}{\includegraphics{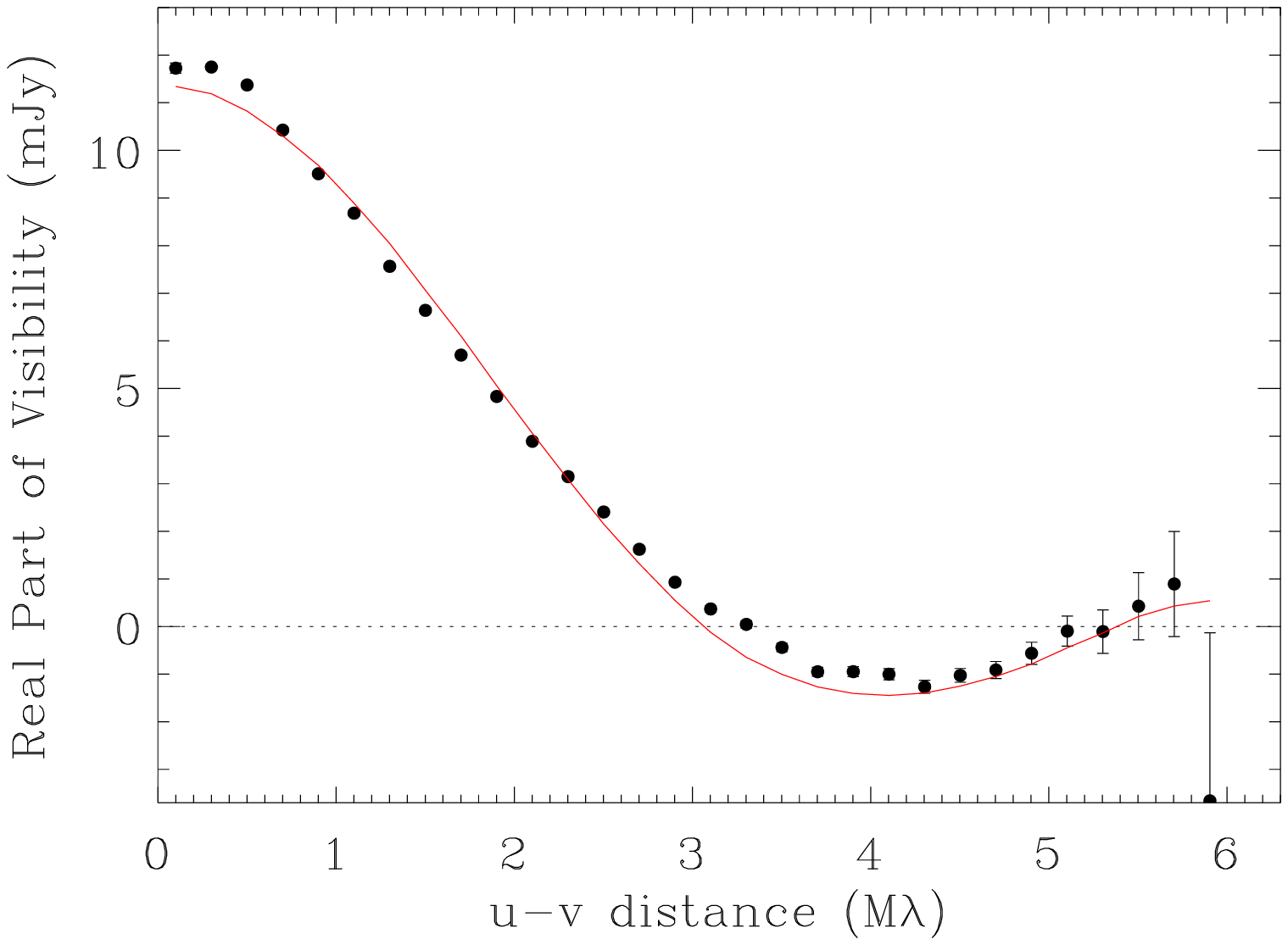}}}
\scalebox{0.4}{\rotatebox{0}{\includegraphics{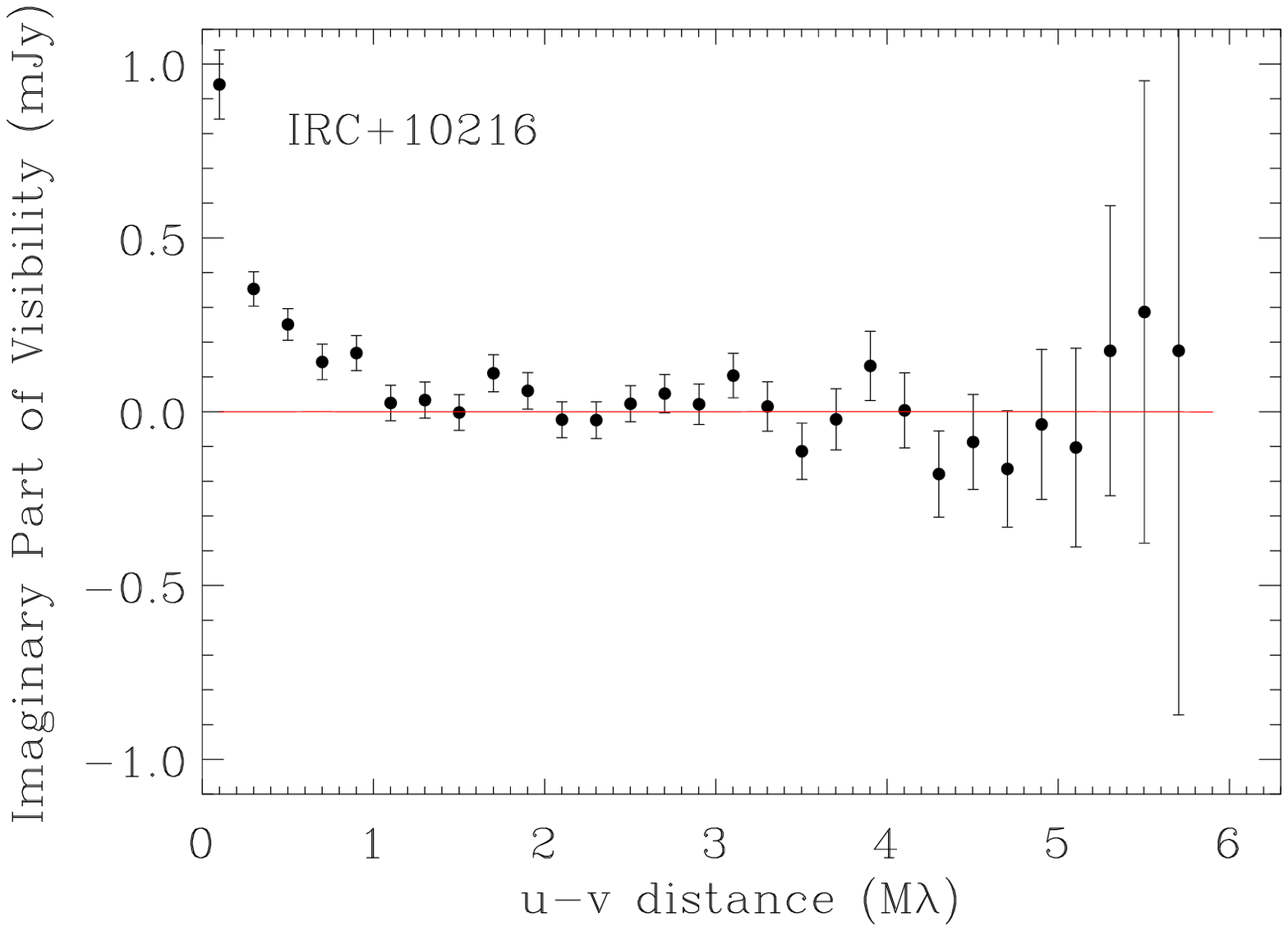}}}
\scalebox{0.4}{\rotatebox{0}{\includegraphics{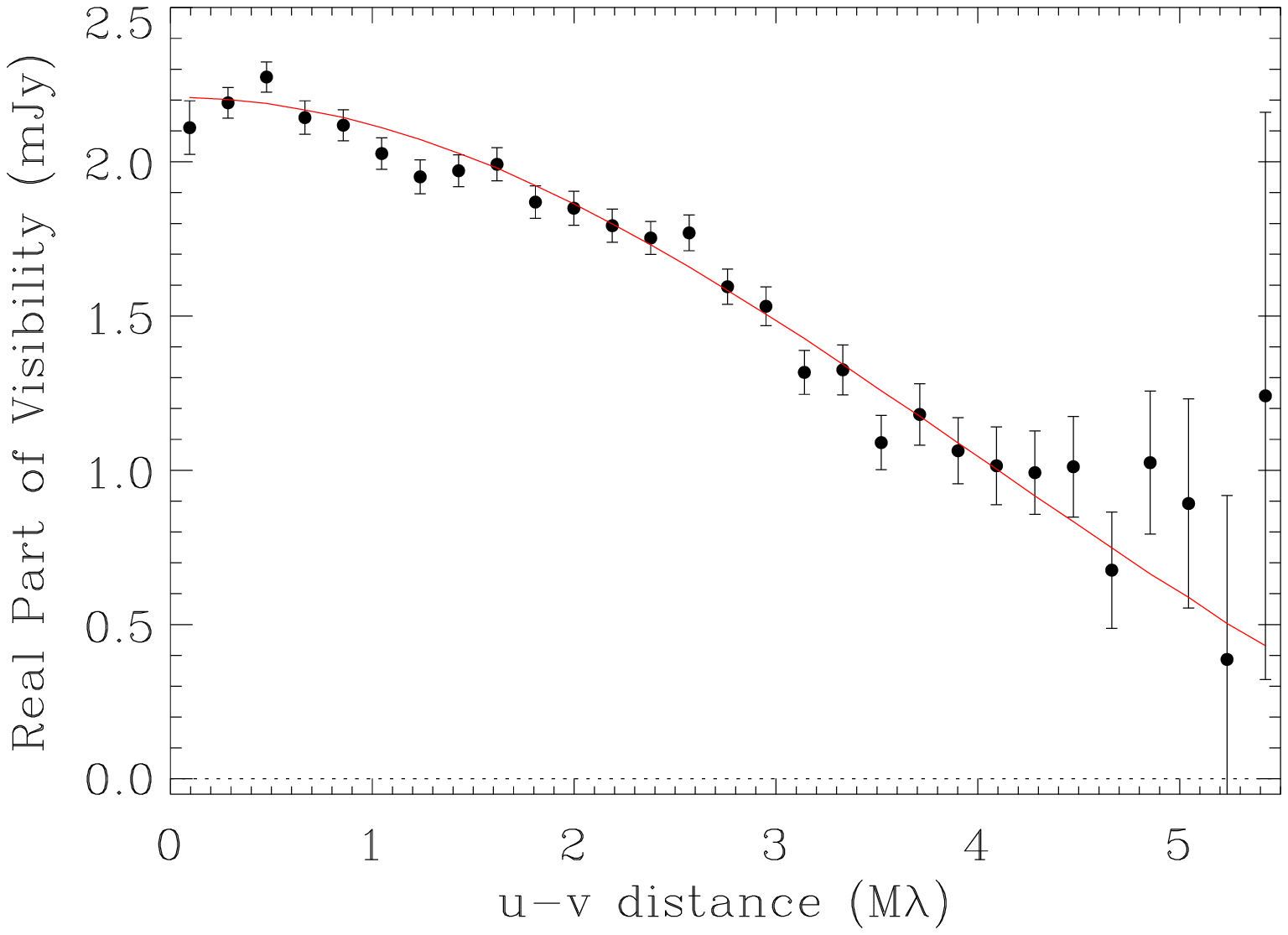}}}
\scalebox{0.4}{\rotatebox{0}{\includegraphics{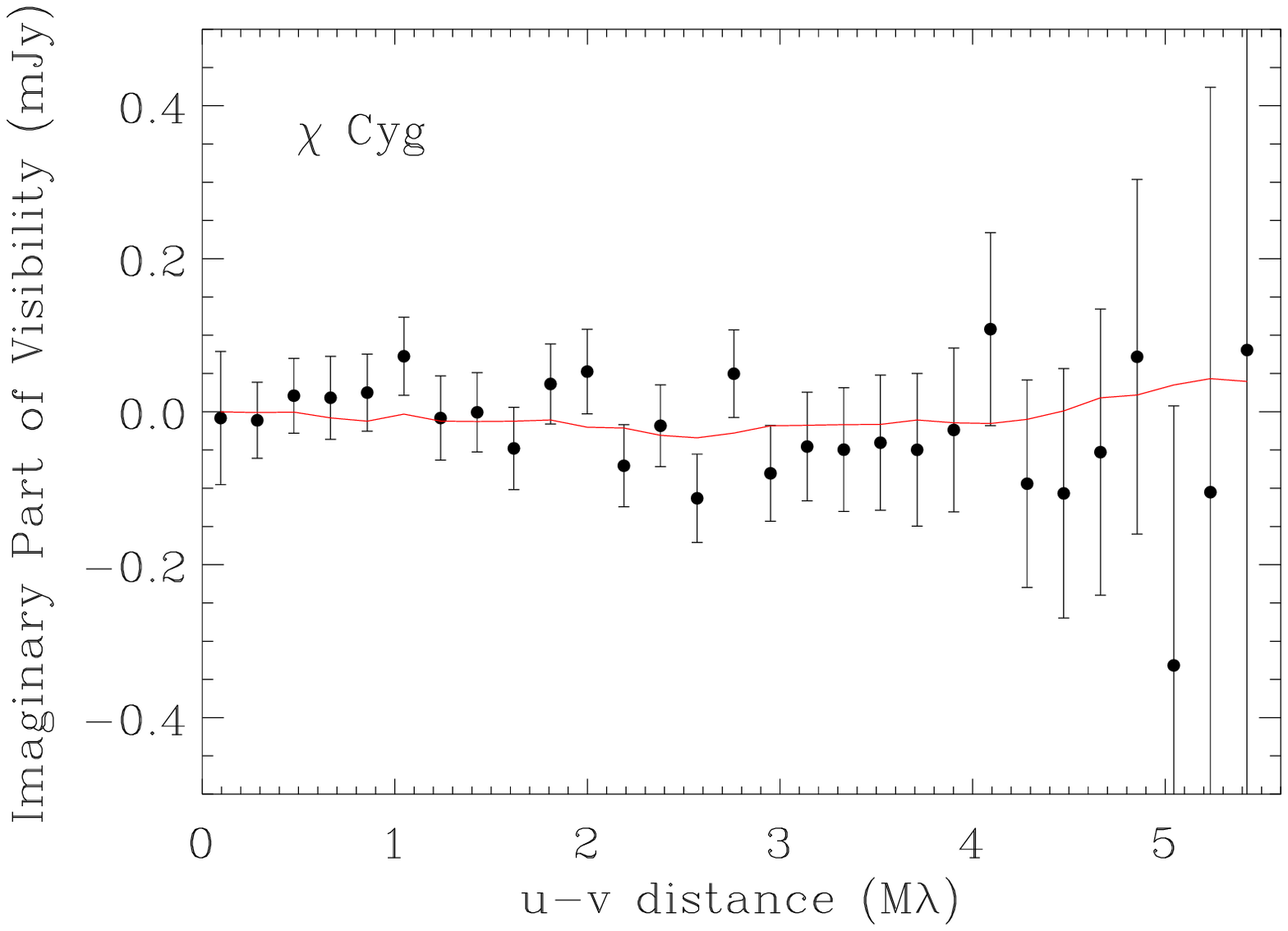}}}

\caption{Real (left) and imaginary  parts (right)
of the visibility amplitude versus baseline length for the four stars in the
  current sample. The solid red lines shows the best-fitting uniform
  elliptical disk models from Table~4.
}
\label{fig:uvplots}
\end{figure*}

\subsubsection{W~Hya}
RM07 previously measured a significant ellipticity of the radio
photosphere of W~Hya ($\sim$0.33). In contrast,
our new data reveal a nearly spherical shape, with only a
slight hint of elongation along the north-south direction.
The mean diameter that we measure for W~Hya from our latest measurements is
$\sim$20\% larger than that found by RM07. The pulsation phase of our
new measurement ($\phi$=0.61) corresponds to the phase where
measurements at optical wavelengths 
have previously shown other AGB stars to exhibit their maximum diameters (Burns et
al. 1998; Young et
al. 2000). However, in the visible band,  this effect is likely the
result of opacity
changes rather than bulk motions of the atmosphere (Young et al.).
Long-term monitoring at radio and/or mm wavelengths will be needed to determine whether
discernible size
changes are linked with the pulsation cycle, or instead occur on
unrelated timescales and
reflect other types of secular changes in the atmosphere (e.g.,
non-radial distortions or large-scale changes in the surface
brightness).

Studies at other wavelengths have found evidence for fluctuating
 degrees of spherical symmetry in the atmosphere of W~Hya. 
For example, for the optical photosphere, Lattanzi et
 al. (1997) reported that the major axis exceeded the minor axis by $\sim$20\%,  although
 no such effect was seen by
Ireland et al. (2004). The mid-infrared measurements by Zhao-Geisler et al. (2015)
suggest a minor-to-major axis ratio of 0.4-0.6 (see also Zhao-Geisler et al. 2011 and
 references therein), and Monnier et al. (2004) also reported deviations
 from a uniform (circular) disk at a wavelength of 2.2$\mu$m.

The position angle of the (slight) elongation of W~Hya measured from our 
VLA observations is nearly  north-south, which agrees to within
uncertainties with the magnetic field axis determined by Szymczak et
al. (1998) from OH maser observations and with the axis along which
Vlemmings et al. (2011) measured a velocity gradient in the SO line at 215~GHz.
However, the earlier measurements of RM07 found the radio photosphere
to be elongated almost east-west, suggesting that its shape and
orientation are variable and not linked to any preferred axis or to
the magnetic field.

 Recently
 Vlemmings et al. (2017) measured the radio photosphere of W~Hya in
 the submillimeter range (0.7~mm) using ALMA and reported a size of
 $(56.5\pm0.1)\times(51.0\pm0.1)$~mas along a PA of
 65.7$\pm$0.3~deg. While it is expected that the size of the radio
 photosphere should be smaller at shorter wavelengths because of
 opacity effects, the position
 angle and degree of flattening are also significantly different from our
 7~mm JVLA measurements. The ALMA data were obtained  at a
 pulsation phase of $\phi\sim$0.3 in 2015 December, approximately 22
 months after our JVLA observation. This suggests that the photospheric
 shape and/or brightness pattern may have changed on these timescales,
 but future contemporaneous observations at different frequencies
 would be useful for ruling out opacity-dependent effects and/or
 changes linked with the stellar pulsation cycle.

\subsubsection{$\chi$~Cyg\protect\label{chisize}}
The S-type star $\chi$~Cyg has not been previously
resolved at radio wavelengths, so we cannot compare its radio size and
shape to previous epochs. However, the 46-GHz flux density
we measure for this star is roughly a factor of two smaller than
predicted by
the radio photosphere model of RM97 (their equation~7),  
and it lies below the
extrapolation of the longer wavelength measurements of RM97 to
$\lambda\sim$7~mm. 
Our current observations were obtained
at an optical phase of 0.75, close to the phase where the bolometric flux
is predicted to be at a minimum (Lacour et al. 2009), compared with
RM97, at the time of whose observations the phase was $\approx$0.09. 

The mean shape of $\chi$~Cyg in our
new observations is found to be nearly spherical (but see also Section~\ref{sparse}),
with a mean diameter
intermediate between the range of values
observed for other M-type Miras
(Table~5; MR07; Matthews et al. 2015). Thus from this single data
point we find no evidence of a significant difference between the
radio sizes of S-type versus M-type Miras. (In contrast, the carbon
star IRC+10216 is nearly twice as large as the M-type Miras; see Section~\ref{IRCsize}).

Previous measurements of the size of $\chi$~Cyg 
at 1.6$\mu$m and a pulsation phase of $\phi$=0.79 were made by 
Lacour et al. (2009), who 
derived a photospheric diameter of 21.49$\pm$0.11~mas
and a size for the molecular layer of 27.35$\pm$0.13~mas. These
authors inferred a temperature of
2032$\pm$32~K for the molecular layer, notably warmer than the
brightness temperature of 1327$\pm$210~K that we derive from our radio
data (Table~5). On the other hand, at 2.2$\mu$m, Perrin et al. (2004) derived a diameter
and temperature for the molecular layer of 30.78$\pm$0.10 and 
1737$\pm$53~K, respectively, at $\phi$=0.76. Part of the apparent discrepancy
likely stems from the fact that the results of Lacour et
al. were derived taking into account limb darkening, whereas Perrin et
al. argued that limb-darkened models do not provide a satisfactory fit
to AGB star data. 

\subsubsection{IRC+10216\protect\label{IRCsize}}
For the carbon star IRC+10216, we find no 
statistically significant changes in the mean diameter, degree
of flattening,
or in the 7~mm flux density compared to previous radio observations
(see Tables~4 and 5
and M12). However, our new
measurements do suggest  a change in the position angle of the 
flattening. 

The radio diameter of IRC+10216 (10.6~AU) is significantly larger
than the three M- and S-type three stars in the sample (which range
from 4.7-7.5~AU; see Table~4). Indeed, its 7~mm radio size is
intermediate between that of M-type AGB stars measured to date and the red supergiant
Betelgeuse ($\sim$17.5~AU, assuming $d$=200~pc; Lim et al. 1998;
O'Gorman et al. 2015). However, the measured radio brightness temperature of
IRC+10216, $T_{\rm eff}=1580\pm260$~K, is significantly cooler  than that of
Betelgeuse at this wavelength, where $T_{\rm eff}\approx$3450$\pm$850~K
(Lim et al.).  

It is presently unclear whether the large radio size of IRC+10216
compared with the M-type sample is a general property of carbon stars
or is instead related to its cool temperature and high luminosity  
(see also Section~\ref{visan}) and/or to its advanced
evolutionary state. Radio opacity is
tightly linked with the ionization of Na and K (RM97), and the
physical radius of carbon stars as measured in the radio may on
average be rather different from oxygen-rich AGB stars owing to the
different physical conditions in their atmospheres. On the other hand,
IRC+10216 is also a highly evolved AGB star that is 
thought to be close to transitioning into a protoplanetary nebula
 (e.g.,
Skinner et al. 1998; Osterbart et al. 2000), and this too may impact
its radio size.  Additional resolved imaging observations of carbon
stars at millimeter wavelengths are needed to address the question of whether IRC+10216
is typical of its class.

Because of its optically
thick and constantly evolving dust envelope, IRC+10216 does not
have a well-defined photosphere at optical and infrared
wavelengths (Stewart et al. 2016). Therefore, in contrast to the M-
and S-type AGB stars that have been resolved, we cannot readily compare
its radio size with a photospheric size measured at other
wavelengths. However, based on the radio diameter $D$ and brightness
temperature $T_{B}$, one may derive a bolometric luminosity for the
star based on the Stefan-Boltzmann relation:
$L_{b}=\pi D^{2}\sigma_{B}T^{4}_{B}$ where $\sigma_{B}$ is the
Stefan-Boltzmann constant. We find $L_{b}\approx7200\pm1300~L_{\odot}$, where
we assume the solar luminosity to be 3.83$\times10^{33}$~erg  s$^{-1}$.
Adopting the stellar effective temperature of 2200~K from Cohen
(1979), we then find a predicted stellar
diameter of $\sim$5.5~AU. This implies that that radio photosphere of
IRC+10216 is roughly twice the size of its ``classical'' photospheric
diameter, the same ratio as seen in M-type AGB stars.

Both the brightness temperature and the  bolometric luminosity 
that we derive for IRC+10216 agree with
the values derived by M12 to within uncertainties. The uncertainty in
$L_{b}$ is, however, quite large and is dominated
by calibration uncertainties in the absolute flux density. 

\subsection{Discussion: the Possible Origins of Secular Shape Changes of Radio Photospheres}
Our new data provide the first compelling evidence that the shapes of
the radio photospheres of AGB stars may not only be non-spherical, but
may evolve  over
time. Previously, Mira was shown to have a non-spherical shape,
with tentative evidence that it had evolved over 14 years
(Matthews et al. 2015). Our new data suggest this phenonemon is likely
to be common, at least for M-type AGB stars. Furthermore, these
results allow us to
exclude several possible explanations for the origin of the
non-spherical shapes of radio photospheres
(see Section~\ref{Intro}), including
stellar rotation, tidal effects, and magnetic fields. In each of these cases,
any induced ellipsoidal shape and/or its axis of 
elongation would be expected to be stable
over timescales of many decades and therefore cannot readily explain the
shape changes observed in a span of $\sim$8-14
years.  

Because we have
just two  epochs of observations separated 
by several years (Table~5), we presently have very limited information on how  rapidly
the shapes and sizes of
radio photosphere may change. Furthermore, although in all cases the observations
in the two epochs were
obtained at different phases of the stellar pulsation cycle, we do not
yet have sufficient information to determine whether changes in
parameters including shape,
radius, or temperature of the radio photosphere are discernible over the course of a single
cycle. To date, our best insight into this question comes from the case
of Mira, where Matthews et al. (2015)
found  that
the ellipticity and position angle 
of the photosphere appeared to be stable on timescales of at least 8
months---a significant fraction of the stellar pulsation period.

Recent 3D hydrodynamic simulations  of AGB star atmospheres by Freytag et al. (2017;
see also Freytag \& H\"ofner 2008) make predictions that
provide a possible framework for the interpretation of our data. In
these models, the dynamics of the atmosphere is governed by the
interaction between long-lived giant convective cells 
(Schwarzchild 1975) with shorter-lived granules and strong
radial pulsations. Such cells may be responsible for the elevation of
material in different parts of the atmosphere and lead to temperature
changes within the radio photosphere. Associated shock waves may also
lead to temperature fluctuations within the radio photosphere. 
Interestingly, even though the aforementioned models adopt spherically symmetric
flows with purely radial, fundamental mode pulsations, the resulting
stars clearly manifest non-spherical shapes and non-radial structures,
with discernible changes in the shape and surface brightness of the
star occurring over time. Specifically, these models predict that the large-scale
convective cells, which are expected to play such an important role in governing the
atmospheric dynamics, should be stable on timescales of order one year,
comparable to the timescale over which the shape of Mira's radio
photosphere appeared to be relatively stable.
Future monitoring
observations with a cadence of a few weeks or less will be important
for 
testing whether this trend is seen in other stars and/or is reproduced
during additional Mira observing epochs.   Meanwhile, further evidence for the
role in giant convective cells in the radio properties of our sample
is discussed below (Section~\ref{nonuniform}). 

Non-radial pulsations are
another potential explanation for the non-spherical shapes of radio
photospheres, although their occurrence in AGB stars is not well-established.
For example,
Stello et al. (2014) have suggested that while  non-radial pulsations
may be present in semi-regular variables, they appear to be absent
from Miras (but cf. Tuthill et al. 1994). 
On the other hand, Freytag \& H\"ofner (2008) have noted
that the distinction between radial and non-radial
pulsations may become blurred, since the type of pulsations present in
their hydrodynamic models combine changes in volume with changes in
the shape of the star---despite initial starting conditions with spherical
symmetry and fundamental mode pulsations.

\subsection{Evidence for Non-Uniform Brightness of the
  Radio Surfaces\protect\label{nonuniform}}

\subsubsection{Visibility Plane Analysis\protect\label{visan}}
Figure~\ref{fig:uvplots} presents another
representation of the $u$-$v$ data, this time showing the 
  azimuthally averaged 
real and imaginary
parts of the visibility amplitude versus projected baseline length for
each of the four  stars in the present sample. The overplotted red lines show the
best-fitting uniform elliptical disk model, based on the parameters
from Table~4. These plots are helpful for highlighting not only
ellipticity, but also deviations in the stellar surface brightness
from a uniform circular or elliptical disk.
Based on Figure~\ref{fig:uvplots}
all four stars show at least some hints of deviations from this simple model,
suggesting asymmetries in shape and/or the presence of brightness
non-uniformities across the stellar surface. For W~Hya and R~Leo, this is most evident in the
imaginary part of the visibilities.
Similar effects have been seen previously on the radio surface of Mira (Matthews et
al. 2015; Vlemmings et al. 2015; Wong et al. 2016) and in recent ALMA
0.7~mm observations of W~Hya (Vlemmings et al. 2017). 

\begin{figure}
\centering
\scalebox{0.4}{\rotatebox{0}{\includegraphics{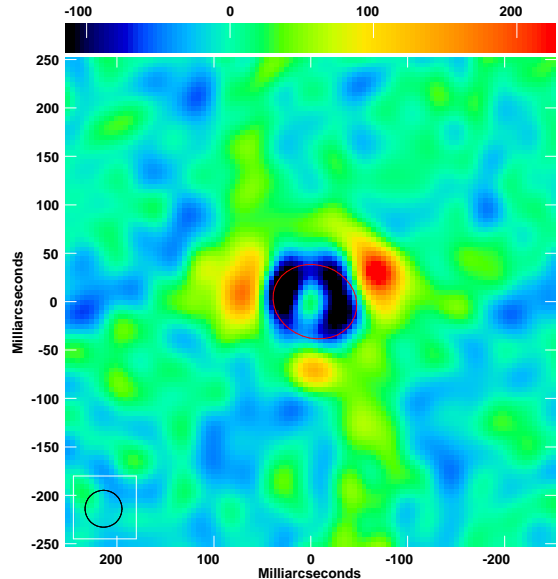}}}

\caption{Residual {\small\sc{CLEAN}} image of IRC+10216, imaged after subtraction of the
 best-fitting uniform elliptical disk fit from the visibility
 data. The size and orientation of the uniform disk fit are
 overplotted as a red ellipse. Intensity units are $\mu$Jy beam$^{-1}$.
 }
\label{fig:IRCresid}
\end{figure}

Based on the visibility plots in Figure~\ref{fig:uvplots},
the most pronounced deviations from a uniform brightness elliptical disk are seen in the case of
IRC+10216, which is also the star in our sample 
with the largest angular extent. To
explore its properties further, we present in
Figure~\ref{fig:IRCresid} a {\small\sc{CLEAN}} image
of the star made after subtraction of the best-fitting uniform
elliptical disk model from the observed visibilities.  The residual
image reveals an oversubtraction across the nominal stellar disk
(deisgnated by a red ellipse),
leaving a clumpy, ring-like distribution of
negative residuals ($\sim$10--12$\sigma$) visible around the
periphery of the star. Outside of this there is a ring of positive
residual emission that extends well beyond the disk radius defined by
the uniform elliptical disk model. These positive residuals include three bright
clumps (with significance 9--17$\sigma$). 
The individual clumps are only 1-2\% of the peak brightness, and
the possibility that they are spurious artifacts cannot yet be
excluded. However, the overall evidence for the 
presence of extended emission appears to be significant and
cannot readily be explained by calibration or imaging artifacts.

To provide a better representation of the brightness distribution of
IRC+10216, we have attempted fits using a variety of more complex models 
to the data. In addition to a uniform elliptical disk, each of these
new models included
one or more additional brightness components, including 
a thin elliptical ring, up to three circular or
elliptical Gaussians, and/or up to three
point sources.  


%
\begin{deluxetable*}{lllllll}
\tabletypesize{\tiny}
\tablewidth{0pc}
\tablenum{6}
\tablecaption{Multi-Component Model Fit Parameters for IRC+10216}
\tablehead{
\colhead{Comp.} &  \colhead{E offset (mas)} & \colhead{N offset
(mas)} & 
\colhead{$\theta_{a}$ (mas)} & \colhead{$\theta_{b}$ (mas)} & \colhead{PA (deg)} &
\colhead{$S_{\nu}$ (mJy)} }

\startdata

Uniform elliptical disk & $-0.18\pm$0.09 & 0.18$\pm$0.08 & 84.4$\pm$0.3 &
71.3$\pm$0.3 & 72.0$\pm$0.4 & 10.62$\pm$0.04 \\

Thin elliptical ring & $-$5.33$\pm$0.83 &  5.90$\pm$0.79 &
146.3$\pm$2.0 & 105.1$\pm$2.1 & 94.3$\pm$1.1 & 1.28$\pm$0.04

\enddata

\tablecomments{Quoted error bars include only the formal fitting uncertainties.}
\end{deluxetable*}

\begin{figure}
\centering
\scalebox{0.4}{\rotatebox{0}{\includegraphics{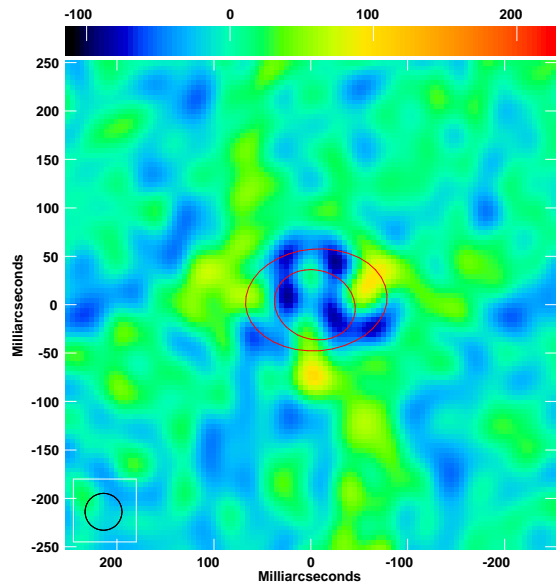}}}

\caption{As in Figure~\ref{fig:IRCresid}, but after subtraction of a model
 comprising a
 uniform elliptical disk plus an elliptical ring  (see Table~6). The
 two components of this model are indicated as red ellipses.
 }
\label{fig:IRCringmodel}
\end{figure}

We find that a significant improvement in the quality
of the fit is achieved by a model that includes just one additional
brightness component, namely a geometrically thin elliptical ring. Parameters of this
model are summarized in Table~6. 
A residual
map, made after subtraction of the model from Table~6 from the IRC+10216
visibility data, is shown in Figure~\ref{fig:IRCringmodel}. 
The
residuals in the map are significantly reduced relative to
Figure~\ref{fig:IRCresid}, although some residual clumps visible across
the disk of the star suggest that the radio surface may be intrinsically
non-uniform. Despite this, we find that 
the addition of one or more components (point sources or Gaussians) to the elliptical
disk+ring model provides only marginal improvement in the
residuals while increasing the number of free parameters. We conclude
that the star likely has a mottled surface, but based on the
present data we cannot uniquely model
it with a small number of brightness components.

For the simple two-component model presented in Table~6, 
all parameters were allowed to vary freely. Despite this,
we find that all of the parameters of the
uniform elliptical disk in this revised model (with the exception of the flux
density) agree to within uncertainties with the original disk-only 
model fit
presented in Table~5. (As we are interested in relative quantities, 
we ignore here the contribution of absolute flux
calibration uncertainty and various systematic effects to the error budget). 

The two-component model that we find to well represent the 
brightness distribution of IRC+10216  is qualitatively
similar to the one recently derived for the red
supergiant Betelgeuse by O'Gorman et al. (2017) using 338~GHz
observations from ALMA, although for Betelgeuse, one circular and one
elliptical Gaussian component
were also included in the published model. For both stars the
``ring'' comprises $\sim$10\% of the total combined flux of the
disk+ring, although in the case
of IRC+10216, the ratio of the major axis diameter of the elliptical ring to the
disk is larger than the Betelgeuse case ($\sim$1.7 compared to
$\sim$1.2). The latter difference may be a consequence of the
different wavelength bands used for the respective measurements, but may also
reflect an intrinsic difference between
the structure and temperature profile of the radio-emitting
atmospheres of the 
two stars.

%
%

We exclude contamination from line emission as a likely 
explanation for the ``ring'' component to the IRC+10216
atmosphere. While 
IRC+10216 is known to exhibit weak, thermally excited (i.e.,
non-masing) 
line emission from various molecules within our observing band, including
HC$_{3}$N
(Chau et al. 2012), the HC$_{3}$N emission is expected to
arise predominately from the circumstellar envelope at distances of hundreds of AU from the star
(Claussen et al. 2011), and we find no evidence for any line
emission detected in our VLA data set. Furthermore, we have
confirmed that evidence for the ``ring'' persists across our entire 8~GHz
observing band. Additional resolved imaging of ultra-high-luminosity red
giants is needed to
test whether this property is common among such stars.

O'Gorman et al. (2017) argued that in the case of Betelgeuse, 
the presence of the ring cannot be attributed to
limb-brightening because it is not located near the limb of the
main disk. A similar argument applies to  IRC+10216.
One possible alternative is that the ring comprises gas elevated
beyond the nominal radio photosphere by giant convective cells (e.g.,
Lim et al. 1998). We also note the models of Freytag et al. (2017)
predict that the coolest and most luminous AGB stars should have
increasingly ill-defined ``surfaces'' (cf. their Figure~8), and the
amorphous and extended appearance of IRC+10216's photosphere at 7~mm wavelengths
qualitatively resembles their model predictions for a star with
$L\sim10^{3}~L_{\odot}$. 

\subsubsection{Results of Sparse Model Imaging\protect\label{sparse}}
The model fits to the visibility data for our sample stars reveal
indications of departures from uniform brightness models
(Section~\ref{visan}). However, because the 
stars are only marginally resolved by the VLA at 7~mm wavelengths, it
is difficult to quantitatively characterize these deviations in the
$u$-$v$ plane, and more
complex models are generally not well-constrained 
(with the exception of IRC+10216; see above). Furthermore, it is 
challenging to identify signatures of possible non-uniformities in the image
plane. This stems from a combination of the relatively low contrast of these
expected features ($\sim$10-20\% contrast relative to the background;
e.g., Freytag et al. 2017; Paladini et al. 2018) and
the need to distinguish real features from artifacts caused by the
limited $u$-$v$ sampling or other limitations inherent to the {\small\sc{CLEAN}}
imaging algorithm (see, e.g., Cornwell, Braun, \& Briggs 1999; Akiyama
et al. 2017a,b).

The {\small\sc{CLEAN}} deconvolved image data presented in Figure~\ref{fig:kntrmaps}, produced
using Briggs  weighting with robustness ${\cal R}$=0, are adequate for
displaying the overall shapes of the stellar isophotes and their
degree of ellipticity, but 
these images do not readily enable identification of possible stellar surface
features (e.g., spots or cool regions) 
or brightness asymmetries. A {\small\sc{CLEAN}} deconvolution using 
uniform weighting (essentially equivalent to Briggs weighting
with ${\cal R}=-5$) can produce images with a slightly improved ($\lsim$20\%)
angular resolution, but  at the cost of higher image noise, reduced
surface brightness sensitivity, and a more
complex dirty beam pattern (Briggs, Schwab, \& Sramek 1999), all of
which degrade the overall image
fidelity.


%
\begin{figure*}
\centering
\scalebox{5.5}{\rotatebox{0}{\includegraphics{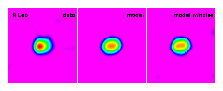}}}
\scalebox{5.5}{\rotatebox{0}{\includegraphics{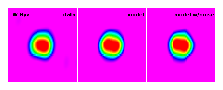}}}
\scalebox{5.5}{\rotatebox{0}{\includegraphics{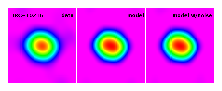}}}
\scalebox{5.5}{\rotatebox{0}{\includegraphics{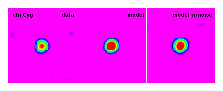}}}
\caption{Sparse model reconstructed images of R~Leo (top
  row), W~Hya (second row),
  IRC+10216 (third row), and $\chi$~Cyg (bottom row) at 7~mm. Compared with
  the {\small\sc{CLEAN}} images shown in Figure~\ref{fig:kntrmaps}, the sparse
  model images
  achieve modest levels of super-resolution, allowing identification
  of additional photospheric features (see Text for details). Each panel shows
  a field-of-view of 190~mas per side. The {\em left-hand}
  panels are images of the stellar data; {\em center} panels
  show noise-free sparse model images of the best-fitting elliptical disk
  models for each star; {\em right-hand} panels show the same models
  with Gaussian noise comparable to the real data. Intensity scales for the data and the models
  are identical. Differences between the data and the smooth and symmetric
  model images
suggest evidence of irregular shapes, asymmetries, and non-uniform brightness of the radio
  photospheres. 
}
\label{fig:sparse}
\end{figure*}

One alternative strategy is to impose a 
modest degree of ``super-resolution'', through the use of a {\small\sc{CLEAN}} 
restoring beam whose FWHM is smaller than
the FWHM of the dirty beam. Following Fish et al. (2016) we define
super-resolution as 
$\lambda/(B\alpha_{r})$, where $B$ is the maximum
baseline length of the interferometer and
$\alpha_{r}\gsim1$. However,  
higher levels of super-resolution in {\small\sc{CLEAN}} ($\alpha_{r}>$1) may
produce spurious clumps and undesirable
image artifacts (e.g., Akiyama et al. 2017a,b), 
particularly in cases where the dirty beam is highly
elliptical (e.g., as is the case for W~Hya) or has significant
sidelobes (Fish et al.).

To circumvent these issues, we employ a new imaging technique
known as {\em sparse modeling} (Honma et al. 2014),
in which images are computed directly by solving the observing equation 
with convex regularization functions (see Honma et al. 2014 and Akiyama et al. 2017a for details). 
In typical radio interferometric data sets, the sampling of
the $u$-$v$ plane is often highly incomplete, making image
reconstruction an under-determined problem (Honma et al.). 
Standard radio interferometric imaging techniques, such as {\small\sc{CLEAN}}, 
compensate by using ``zero padding''
to replace unsampled grid points in the $u$-$v$ plane and to infer the 
solution using the inverse Fourier transformed image (the so-called 
``dirty residual map''). However, this results in
degradation of the shape of the beam and an increase in sidelobe levels. 
The sparse modeling approach instead can derive a unique solution
from an infinite number of possible images by utilizing a convex sparse regularization function.
The latest algorithm utilizes two convex regularization function of
the brightness 
distribution, its $\ell_1$-norm and Total Variation (TV) [or the
improved variant
Total Squared Variation (TSV)]. These regularizers penalize sparsity in the 
brightness distribution and its gradient, respectively 
(Akiyama et al. 2017a,b; Kuramochi et al. 2018). This technique can 
be used to achieve resolutions as high as $\sim$30\% of the
diffraction 
limit while maintaining image fidelity.

Using the latest sparse modeling code implementation of Kuramochi et al. (2018),
we reconstructed images using $\ell_1$+TSV regularization (Kuramochi
et al.) and the
MFISTA algorithm described in Akiyama et al. (2017a). We computed a
grid of 16 
images for each of our sample stars, spanning a range of four values for the regularization parameters
${\bar\eta}_{l}$=[10, 1, 0.1, 0.01] for $\ell_1$-norm and
${\bar\eta}_{t}$=[1e6, 1e5, 1e4, 1e3] for TSV, respectively (see Kuramochi et
al. for a definition of these parameters). We adopted a field-of-view
of 260~mas with a 100$\times$100 pixel gridding. 
For each star, we also created additional grids of images from model
$u$-$v$ data representing the best-fitting uniform elliptical disk
model to the visibility data (see Table~4), both with and without the addition of
realistic Gaussian noise. These model data were generated
using the AIPS tasks {\small\sc{OMFIT}} and {\small\sc{UVMOD}}. 

To determine the best image, we use a combination of the ``leave-one-out''
cross-validation errors (LOOE; Obuchi \& Kabashima 2016; Obuchi et
al. 2017) and a visual comparison between the
sparse model stellar images and the corresponding sparse model
elliptical disk images for each value of ${\bar\eta}_{l}$ and
${\bar\eta}_{t}$. When LOOE is large, the data
and the elliptical disk models are virtually indistinguishable (i.e., the
images are not significantly super-resolved). On the other hand,
attempting to over-resolve the data can lead to spurious image artifacts despite
small LOOE values. However, because these artifacts tend to appear nearly identical in both
the real images and the uniform elliptical disk images, such images
can be readily rejected as providing a poor representation of the true
source brightness distribution.  

Based on the application of the above criteria, we present in Figure~\ref{fig:sparse}
our best sparse model images for each of the four program stars.
In each case, we also show the corresponding uniform elliptical
disk model image, with and without noise. For R~Leo, our images appear to reveal a clear
asymmetry in the radio photosphere. W~Hya and $\chi$~Cyg also exhibit
subtle deviations from a purely symmetric shape, and for 
$\chi$~Cyg we see that the uniform elliptical disk fit
over-predicts the intensity near the center of the star (see also
Figure~\ref{fig:uvcuts}). For IRC+10216, a uniform elliptical disk
model predicts a flatter intensity gradient across the star than
observed, consistent
with the negative residual seen in the IRC+10216 {\small\sc{CLEAN}} image shown in
Figure~\ref{fig:IRCresid} and with the radial intensity cuts shown in Figure~\ref{fig:uvcuts}).

The degree of super-resolution that we are able to achieve for each
star varies depending on the SNR as well as the $u$-$v$ coverage and
other factors that will be explored in future work. 
For IRC+10216, we obtain only a marginal
improvement in resolution over {\small\sc{CLEAN}} ($\sim$97\% of the
diffraction limit), but results are
significantly better for R~Leo, W~Hya, and $\chi$~Cyg, where we
achieve images with $\sim$75\%, 60\%, and 55\% of the diffraction
limit, respectively. 
These preliminary results suggest that sparse model imaging algorithm
appears to be a promising tool for aiding in the interpretation of stellar imaging data at radio
and millimeter wavelengths.

\section{Summary}
We have used the JVLA to image the radio photospheres of four nearby,
long-period variable stars at 7~mm wavelengths. Through fits to the
visibility data we find that all four stars are clearly
resolved and exhibit shapes that range from nearly spherical to an
ellipticity of $\sim$0.17. 

A comparison to measurements taken during
previous observational epochs several years earlier 
shows that in all cases some of the photospheric properties (mean diameter,
shape, orientation, and/or flux density) appear to have changed with time. These
secular changes help  to rule out several possible causes for the
non-spherical shapes of radio photospheres, including tidal effects,
magnetic fields, or binary companions. Instead, the most probable
explanation for these shapes appears to be manifestations of large-scale convective
flows and/or pulsation.

We have shown that the sparse modeling imaging technique provides a
means to achieve a modest degree of super-resolution in the images
of radio photospheres obtained with the VLA. 
Sparse model images for the four program stars
provide further 
evidence for irregular photospheric shapes and non-uniform brightnesses across the
radio surfaces.
Sparse model imaging thus appears to be a promising  new tool for 
aiding the interpretation of stellar imaging data at radio and
(sub)millimeter wavelengths. 

The radio photosphere of the carbon star IRC+10216 has a diameter
nearly twice as large
as that of the other three M- and S-type stars in our sample. Based on fits to the
real and imaginary parts of the visibilities, IRC+10216 also exhibits
a pronounced deviation from a pure uniform elliptical disk. The data
can be well-fitted by a combination of a uniform elliptical disk plus
a more extended component (modeled as a 
ring-like structure) comprising $\sim$10\% of the total
flux. 

As described in the Appendix, 
our JVLA observations of IRC+10216 also permit a new measurement of the
star's proper motion. Our results are in agreement with previous
values obtained from radio wavelength measurements and do not support
recent claims of observable astrometric signatures from a binary companion.

\acknowledgements
The observations presented here were part of NRAO programs 14A-026, AR446, and AM845. L.D.M. 
gratefully acknowledges guidance from E. Waagen in the use of AAVSO
resources, as well as support from award 1516106 from the National Science Foundation.
K.A. is supported
by a Jansky Fellowship from 
the National Radio Astronomy Observatory. 
\appendix
\section{The Absolute Position and Proper Motion of IRC+10216}
As described in Section~\ref{nomasercal}, the use of self-calibration
on the stellar line emission to improve calibration of 
the complex gains precludes measuring the absolute position of
three of our target stars. However, the calibration technique adopted
for IRC+10216 preserves the absolute positional information, enabling
a measurement of the proper motion of the star.

To measure the absolute position of IRC+10216, we produced a {\small\sc{CLEAN}} image of
the star with the same parameters as given in Table~4, but without applying any a
priori positional shifts. This image was then fitted with a
single-component Gaussian using the AIPS task {\small\sc{JMFIT}}, yielding
the position quoted in Table~1. Following M12, we estimate uncertainties
of 10~mas in both the RA and DEC coordinates. 

The first constraints on the proper motion of IRC+10216 were obtained
by Becklin et al. (1969), who obtained upper limits of 30~mas
yr$^{-1}$ by using a comparison between their own measurement on a
red-sensitive photographic plate (epoch 1969.274) with another from a
1954 {\it Palomar Sky Survey} plate. Using radio wavelength
measurements, Menten et al. (2006) and M12 subsequently 
confirmed a proper motion of the star toward a northeasterly
direction.

More recently, Sozzetti et al. (2017) reported astrometric
measurements of IRC+10216 based on archival (1995-2001) $I$-band
($\sim$800~nm) data. They found an east-west component of the proper
motion, $\mu_{x}$, 
comparable to within uncertainties with that
derived by M12, but reported a significantly different north-south
component, $\mu_{y}$, and postulated that this difference may be due
to the effects of a binary companion and its so-called
variability-induced motion (VIM)  on the apparent motion of the star.

\begin{figure}
\centering
\scalebox{0.53}{\rotatebox{0}{\includegraphics{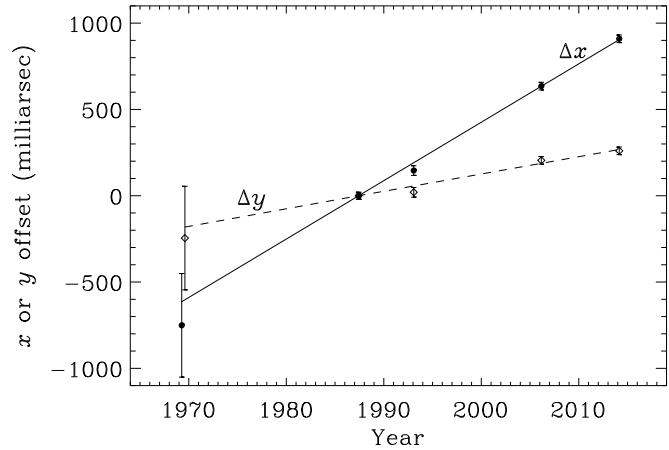}}}

\caption{Proper motion of IRC+10216 as measured from the current JVLA
  observations (epoch 2014.145) and four previous epochs (1969.274,
  1987.419, 1993.07, and 2006.145) taken from Becklin et al. (1969),
  Menten et al. (2006), and M12. The $x$ (east-west) offsets are
  plotted with filled circular symbols and the $y$ offsets (north-south) as
  diamonds. Points from the 1969.274 epoch are
  offset slightly in time for clarity. Epoch 1987.419 is adopted as the
  reference position. Positive offsets indicate
  motions to the east ($x$) or north ($y$).
The solid and dashed lines are weighted least-squares fits
  to the $x$ and $y$ offsets, respectively.  See Appendix for discussion.  }
\label{fig:PM}
\end{figure}

Figure~\ref{fig:PM} presents the previous proper motion measurements
of IRC+10216 from Becklin et al. (1969), Menten et al. (2006), and
M12, along with a measurement from our current data. We adopt the
convention that offsets to the east and to the north are positive, and
following M12, we take 1987.419 as the reference epoch. The
lines on the figure represent weighted least-squares fits to the $x$
and $y$ offsets, respectively, of the combined data. From this we find
a proper motion of
($\mu_{x}$, $\mu_{y}$)=(33.84$\pm$0.7, 10.0$\pm$0.7)~mas
yr$^{-1}$. These values
agree with the determination of M12 to within uncertainties. Our
$\mu_{x}$ determination also agrees with Sozzetti et al. (2017), but
we find significant disagreement in the $y$ component of motion, where
the latter authors report $\mu_{y}$=30.22$\pm$2.02 mas yr$^{-1}$ based
on a single-star fit to the data and $\mu_{y}$=25.43$\pm$1.69~mas yr$^{-1}$ based
on a ``VIM+acceleration'' model, intended to account from perturbations
from a suspected binary companion. A possible explanation for this
discrepancy in the $y$ component of motion is that the 800~nm data
analyzed by Sozzetti et al. are
not sampling the stellar photosphere, but instead material in the
circumstellar envelope whose emission morphology observed at
optical and infrared wavelengths is known to exhibit temporal changes
unrelated to orbital motions (Osterbart et al. 2000; Stewart et al. 2016).

The time elapsed between our recent measurement and the previous
high-precision radio measurement of M12 ($\sim$8~yr) is slightly
larger than the time span of the measurements presented by Sozzetti et
al. (2017). However, we find no evidence for excursions of the motion of
the star compared  with an extrapolation of the data between
1969-2006. Radio measurements to date therefore show no  
evidence for wobble caused by a companion with a significance greater
than $\sim1.5\sigma$, or $>$1~mas yr$^{-1}$.


%

\end{document}